\documentclass[fontsize=12pt,a4paper,DIV12]{scrartcl}

\usepackage[utf8]{inputenc}
\usepackage{graphicx}
\usepackage{authblk}
\usepackage{amsmath}
\usepackage{amsthm}
\usepackage{amssymb}
\usepackage{enumitem}
\usepackage{url}

\usepackage{subcaption}

\graphicspath{{Figures/}{imgs/}}

\title{
  Approximating the Bundled Crossing Number\thanks{ This work was initiated
during the Workshop on Geometric Graphs
in November 2019 in Strobl, Austria. We would like to thank Oswin Aichholzer, 
Fabian Klute, Man-Kwun Chiu, Martin Balko, Pavel Valtr for their avid discussions during the workshop.
The first author has received funding from the European Union’s Horizon 2020 research and innovation programme
under the Marie Skłodowska-Curie grant agreement No 754411.
The second author has been supported by the German Research Foundation DFG Project FE~340/12-1.}
}

\author[*]{Alan Arroyo
}
\author[**]{Stefan Felsner
}  

\affil[*]{IST Austria, Klosterneuburg, Austria\\
\texttt{alanarroyoguevara@gmail.com}}
\affil[**]{Institut für Mathematik, Technische Universität Berlin, Germany
\texttt{felsner@math.tu-berlin.de}}

\renewcommand\Authands{ and }
\date{\today}

\newtheorem{theorem}{Theorem}

\newtheorem{proposition}{Proposition}
\newtheorem{lemma}{Lemma}

\newtheorem{observation}{Observation}
\newtheorem{definition}{Definition}

\newtheorem{remark}{Remark}

\def\bcn{\text{bc}}
\def\crn{\text{cr}}
\def\SS{\mathcal{S}}
\def\BB{\mathcal{B}}
\def\EE{\mathcal{E}}

\def\NN{\mathcal{N}}

\def\term#1{{\em #1}\marginpar{\raggedright\textit{\small #1}}}
\def\tem#1{{\em #1}}

\begin{document}
\maketitle

\begin{abstract}
   Bundling crossings is a strategy which can enhance the readability of
  drawings.  In this paper we consider good drawings, i.e., we require that
  any two edges have at most one common point which can be a common vertex or
  a crossing. Our main result is that there is a polynomial time algorithm to
  compute an 8-approximation of the bundled crossing number of a good drawing
  (up to adding a term depending on the facial structure of the drawing).  In
  the special case of circular drawings the approximation factor is 8 (no
  extra term), this improves upon the 10-approximation of Fink et
  al.~\cite{FHSV16}.  Our approach also works with the same
  approximation factor for families of pseudosegments, i.e., curves
  intersecting at most once.  We also show how to compute a
  $\frac{9}{2}$-approximation when the intersection graph of the
  pseudosegments is bipartite.
\end{abstract}

\section{Introduction}
The study of bundled crossings is a promising topic in Graph Drawing due to
its practical applications in Network Visualization and the rich connections
with related areas such as Topological Graph Theory. One of the mantras
motivating the study of crossing numbers is that ``reducing crossings can
improve the readability of a drawing, leading to better representation of
graphs". The study of bundled crossings provide an alternative way to assess
readability by allowing crossings of a drawing to be bundled into regular
grid-patterns with the goal of minimizing the number of bundles instead of
minimizing the number of individual crossings.

\begin{figure}[thb]
    \centering
    \includegraphics[width=0.85\textwidth]{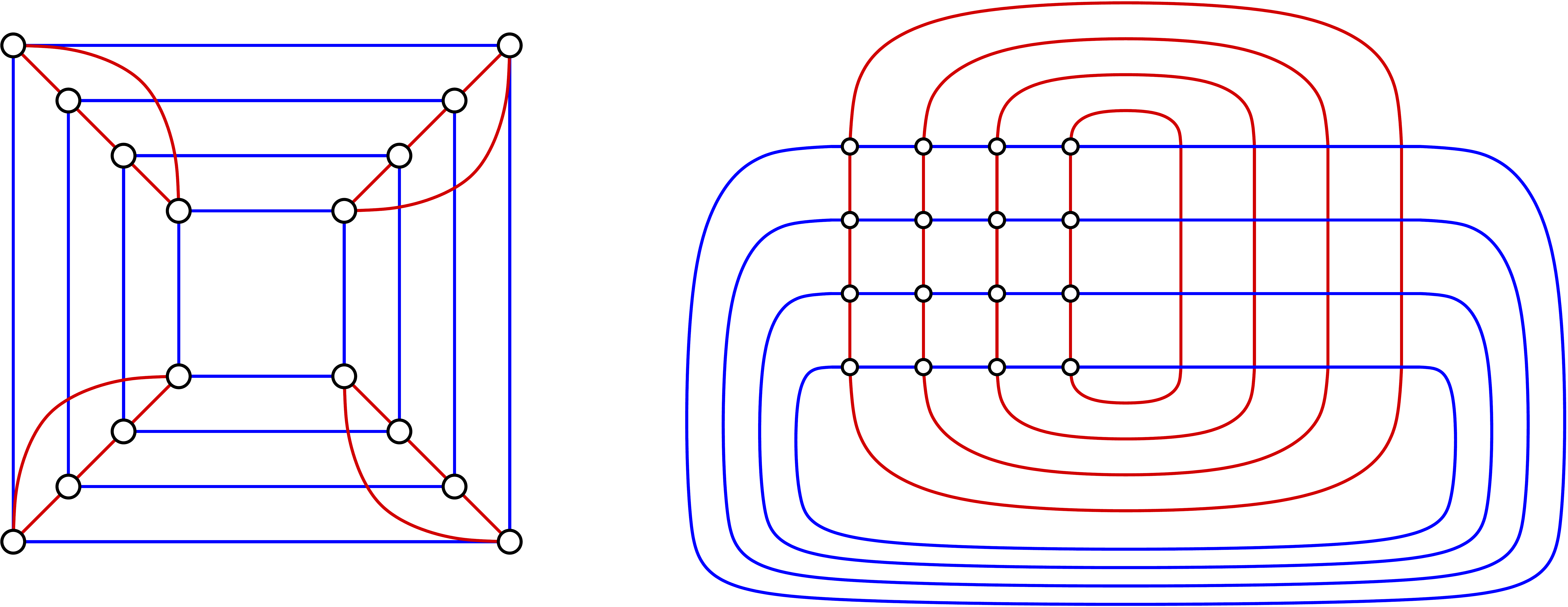}
         \caption{The crossing number of $C_4\Box C_4$ is 8 (left)
    but there is a drawing with 16 crossings which can be viewed as a
    single bundled crossing (right).}
         \label{fig:c-product}
\end{figure}

The \term{crossing number} of a graph $G$ is the minimum integer $\crn(G)$ for
which $G$ has a drawing in the sphere with $\crn(G)$ crossings. Computing the
crossing number of a graph is a notoriously hard problem. There are long
standing conjectures regarding the crossing numbers of complete graphs
\cite{harary1963number,beineke2010early} and complete bipartite graphs
\cite{turan1977welcome}.  Another family whose crossing numbers have been
intensely studied are cartesian products of two cycles $C_m\Box C_n$. It is
conjectured that their crossing number is $\crn(C_m\Box C_n)=(m-2)n$ for,
$3\leq m \leq n$ \cite{harary1973toroidal}. Figure~\ref{fig:c-product}(left)
indicates how to draw products of cycles with few crossings, these drawings
show $\crn(C_m\Box C_n)\leq (m-2)n$.

To define bundled crossings, we consider the \tem{planarization} of a drawing
$D$ in the sphere, this is the plane graph obtained by replacing crossings by
degree-4 vertices (we always assume that a crossing point belongs to only two
edges). A \tem{bundled crossing} or \term{bundle} of $D$ is a subgraph of the
planarization of $D$ isomorphic to an $n\times m$-grid graph ($n$, $m\geq 1$),
whose vertices are exclusively crossings.  A drawing of $C_4\Box C_4$ where
all crossings can be assigned to a single bundle is shown in
Figure~\ref{fig:c-product}(right).

A \tem{bundling} of $D$ is a partition of the crossings of $D$ into disjoint
bundles.  The \term{bundled crossing number} $\bcn(D)$ of a drawing $D$ is the
minimum number of bundles in any bundling of $D$, whereas the bundled crossing
number $\bcn(G)$ of a graph $G$ is the minimum $\bcn(D)$ taken over all
drawings $D$ of $G$. From the fact that $C_4\Box C_4$ is not planar and
Figure~\ref{fig:c-product}, it follows that $\bcn(C_4\Box C_4)=1$.  Indeed,
$\bcn(C_m\Box C_n)=1$ for $n,m\geq 3$.

\subsection{Previous work}

Schaefer in his survey on crossing number problems~\cite{survey1} suggests to
also consider bundlings of crossings. Alam et al.~\cite{AFP16} were the first
to consider the problem from a graph drawing viewpoint.  Later, Fink and
coauthors considered the problem of computing $\bcn(\cdot)$ both in the
\term{free-drawing} variant, when a graph $G$ is the input, and the goal is to
compute $\bcn(G)$, and in the \tem{fixed-drawing} variant, where the drawing
$D$ is the input, and the goal is to find $\bcn(D)$, i.e., to assign the
crossings to as few bundles as possible. In this work we will focus on the
fixed-drawing variant\footnote{In this work we are interested in
  bundling connected drawings, hence, faces of a bundle bounded by
  "squares" are empty. In previous literature, the emptyness of squares was
  part of the definition of bundled crossings.}.

Fink et al.~\cite{FHSV16} showed that computing $\bcn(\cdot)$ is NP-hard in
the fixed-drawing variant of the problem. The hardness of the free-drawing
variant has been shown by Chaplick et al.~\cite{CD+19}. An algorithm that
computes a 10-approximation of $\bcn(D)$ for circular drawings was presented
by Fink et al.~\cite{FHSV16}, here a \term{circular drawing} of a graph is one
where vertices are drawn on a circle and edges are drawn inside the
circle. Circular drawings are assumed to be simple, where a \term{simple
  drawing} is a drawing in which (1)~any two edges intersect at most once;
(2)~the intersection between any two edges is either a common end or a
crossing; and (3)~no three edges share a crossing.  Our work departs from the
question of whether $\bcn(D)$ can be approximated on classes of drawings
beyond the class of circular drawings.
 
\subsubsection{More related work.}
Bundling the edges of a graph is a commonly used technique in the
visualization of large and dense graphs or networks.  Together with van Wijk,
Holten introduced the idea of edge bundling to the area of information
visualization~\cite{Holten06,HoltenW09}.  Subsequently the concept was studied
by many researchers, see for
example~\cite{GansnerHNS11,LambertBA10,PupyrevNBH16,CuiZQWL08,ErsoyHPCT11,BRHMD17,ZPG19,GK06}.
In contrast to the concept of the bundled crossing number most algorithms used
in information visualization for edge bundling are of heuristic nature and do
not provide any guarantees on the quality of the solution.
 
A related definition to bundled crossings are \emph{block crossings}.  This
notion was studied in the context of layouts of metro maps~\cite{FinkPW15} and
storyline visualizations~\cite{DF+17}, but is less general than the later
devised idea of bundled crossings.  Another approach to formalize the idea of
edge bundling are so-called \emph{confluent drawings}.  Here, edges are drawn
as continuous curves that are allowed to merge and split similar to switches
in a train network.  These drawings were introduced by Dickerson et
al.~\cite{DEGM05} and subsequently studied by the graph drawing community%
~\cite{EppsteinS13,EppGMHN05,EppHLNSV16,FGKN19,HuiPSS07}.  Recently, these
drawings were also picked up in more applied contexts, using heuristic
approaches~\cite{BRHMD17,ZPG19}.

\subsection{Our contribution}
\label{subsec:contribution}

To prepare for the statement of our main results we first show how to reduce
the problem of computing $\text{bc}(\cdot)$ for a graph drawing $D$ to computing
$\text{bc}(\cdot)$  for a \term{set of strings}. Secondly, we introduce a
special configuration called a toothed-face which plays a special role in
this work.

To each graph drawing $D$ we associate a set $\EE$ of strings obtained in two
steps: first, delete all the uncrossed edges of $D$; second, for each edge $e$
of $D$, slightly remove two small bits of~$e$ including its endpoints to
obtain a \term{string} (a closed arc). The obtained set $\mathcal{E}$ of
strings is the drawing of a matching, thus $\text{bc}(\EE)$ is
well-defined. Moreover, the bundlings of $D$ are in one-to-one correspondence
with the bundlings of $\EE$, so often in this work we restrict ourselves to
study bundlings of sets of strings.

In connection with the crossing number minimization problem it is natural to restrict
the attention to simple drawings, in this context they are often referred to as
\emph{good drawings}. The set of strings associated with a good drawing turns
out to be a \term{family of pseudosegments}, i.e., it is a set of strings with
the property that no string self-intersects and no two strings cross more than
once.

\begin{figure}[th]
    \centering
    \includegraphics[width=0.55\textwidth]{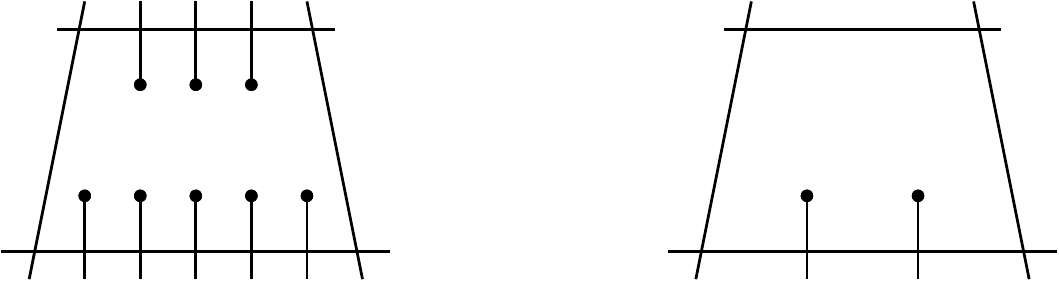}
         \caption{Examples of toothed-faces.}
         \label{fig:toothed_face}
\end{figure}

Any set of strings divides the plane into open regions called {\em faces}. A
string {\em ends} in a face $F$ if one of its endpoints is incident with the
boundary of $F$. Before defining toothed-faces, keep in mind that the boundary
of a face is not necessarily the same as the boundary of its closure. For
instance,  Figure \ref{fig:toothed_face} shows two examples of faces where 
their boundaries include pieces of strings ending in the face while their closure is bounded by only four pieces of strings.


A \term{toothed-face} is a face $F$ of a set of strings such that (1) at least
one string ends in $F$; (2) the closure $\overline{F}$ is bounded by exactly
four pieces of strings; and (3) all the crossings between the strings ending
in $F$ and the boundary of $\overline{F}$ occur in only two opposite
string-pieces among the four string-pieces bounding $\overline{F}$.  For each set of strings
$\EE$ we will let $t(\EE)$ be the number of toothed-faces. If $\EE$ is
associated to a drawing $D$, then we let $t(D):=t(\EE)$. The next is our first
main result.

\begin{theorem}\label{thm:main_8approx}
  For a connected good drawing $D$ with $t=t(D)$ there is a polynomial-time
  algorithm to compute a bundling with at most $8 \text{bc}(D)+t$ bundles.
\end{theorem}

The algorithm used in Theorem \ref{thm:main_8approx} is a simple greedy
approach that has been known for some time.  In the next section we will
informally explain this greedy algorithm. 

Our main contribution is to improve Fink et al. 10-approximation on two
fronts. First, since circular drawings do not induce toothed-faces,
Theorem \ref{thm:main_8approx} shows that the greedy approach produces an
8-approximation instead of a 10-approximation for circular drawings. Second,
Theorem \ref{thm:main_8approx} extends the guarantee of obtaining an
approximate solution to the more general class of simple drawings.

In our second main result we improve the approximation factor when the drawing
$D$ is as in Thm.~\ref{thm:main_8approx} and has the additional property that
the intersection graph of the edges of~$D$ is bipartite. In such case we
call~$D$ a \term{bipartite instance}.

\begin{theorem}\label{thm:approx_bipartite}
  If $D$ is a bipartite instance with $t=t(\EE)$ toothed-faces, then there is a
  polynomial-time algorithm to compute a bundling with at most
  $\frac{9}{2}\text{bc}(D)+\frac{1}{2}t$ bundles.
\end{theorem}

\subsection{An easy example: Bundling bi-laminar families of chords}

In this section, in an informal approach, we consider a concrete example that
captures many of the concepts that will be used in the later parts of the
paper. The approximation algorithm of Fink et al.~\cite{FHSV16} is based on
the very same concepts that are introduced here.

\begin{figure}[ht]
    \centering
    \begin{subfigure}[b]{0.19\textwidth}
         \centering
         \includegraphics[width=\textwidth]{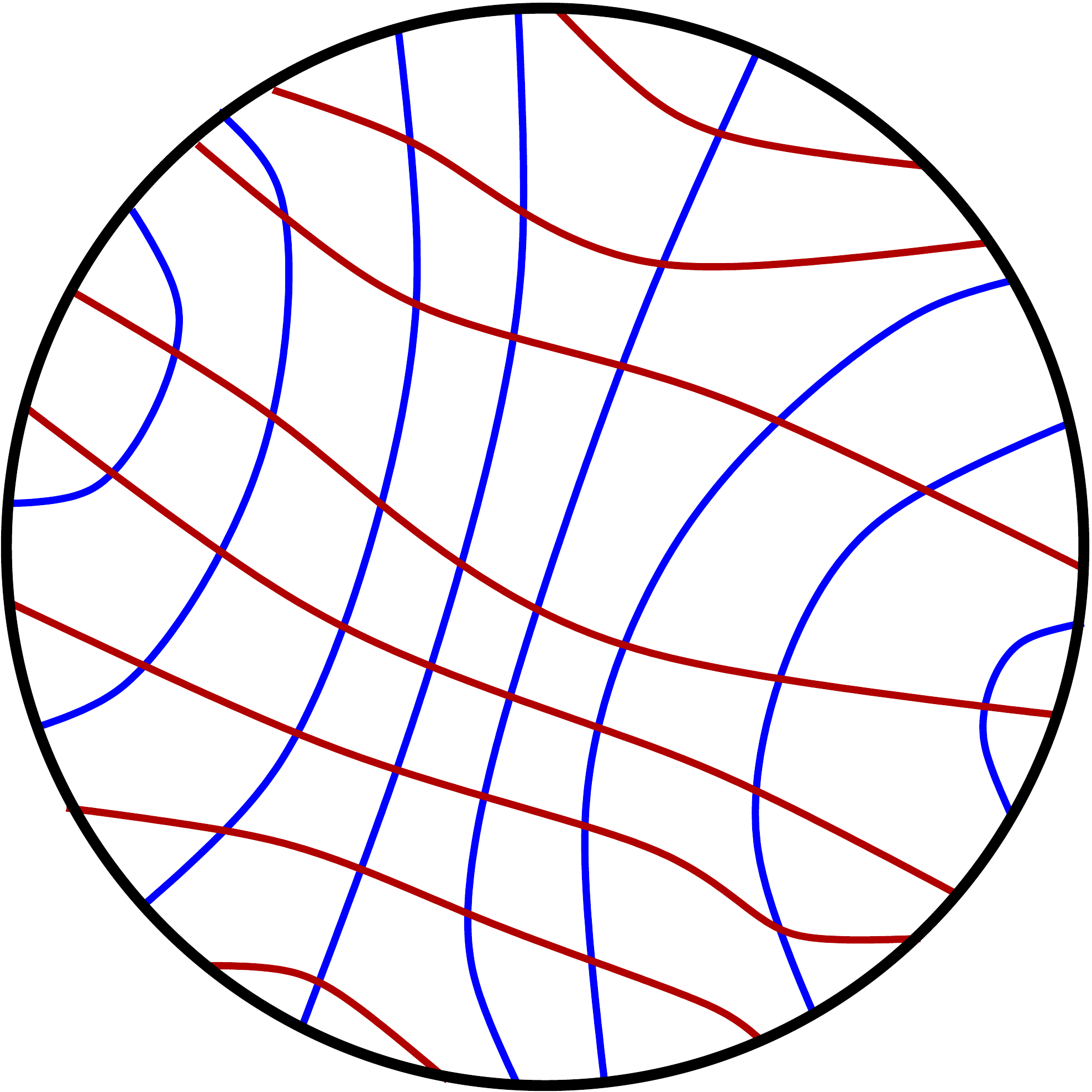}
         \caption{}
         \label{subfig:laminar}
     \end{subfigure}\hskip1cm
    \begin{subfigure}[b]{0.19\textwidth}
         \centering
         \includegraphics[width=\textwidth]{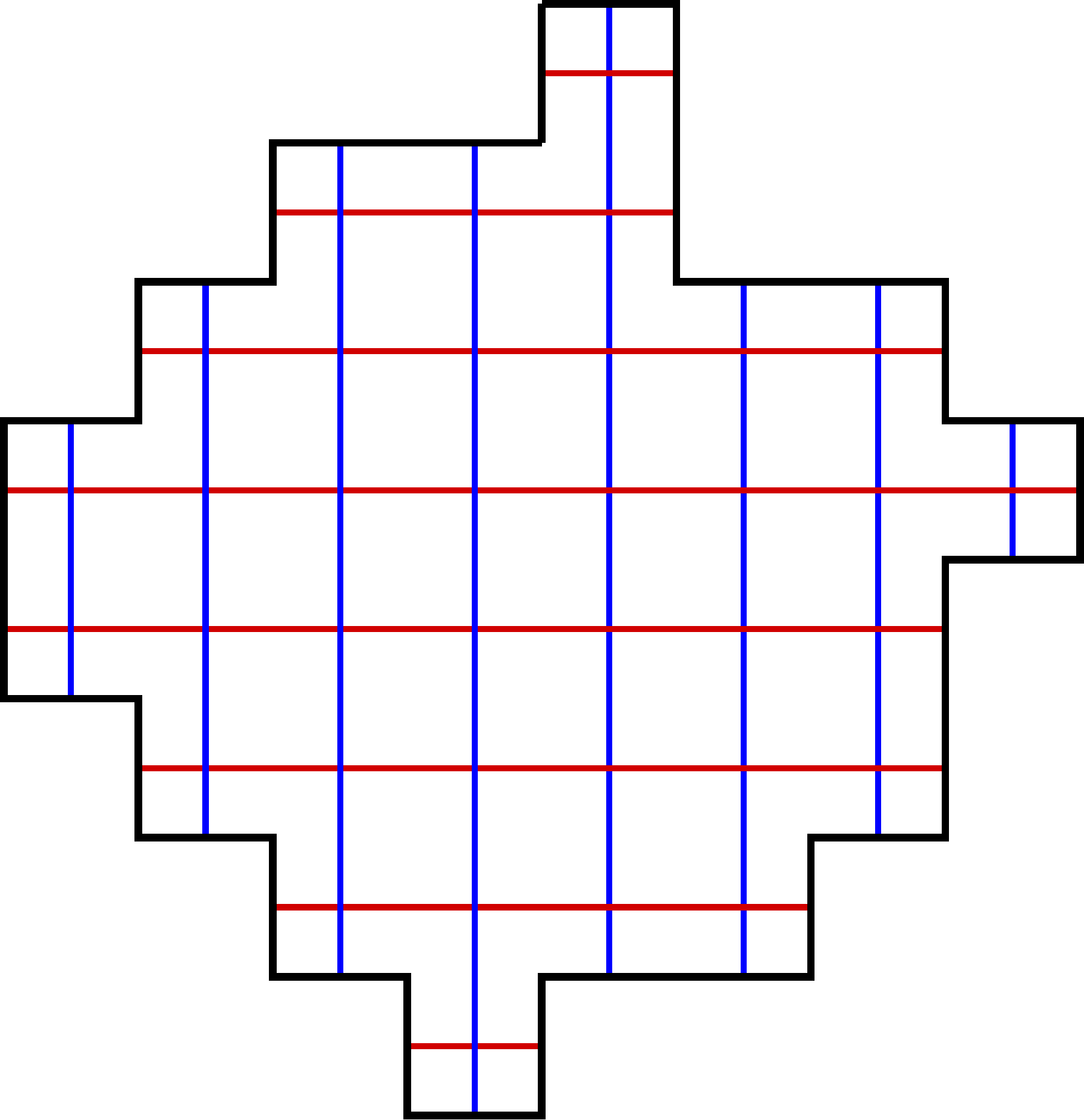}
         \caption{}
         \label{subfig:laminar_orth}
     \end{subfigure}\hskip1cm
    \begin{subfigure}[b]{0.19\textwidth}
         \centering
         \includegraphics[width=\textwidth]{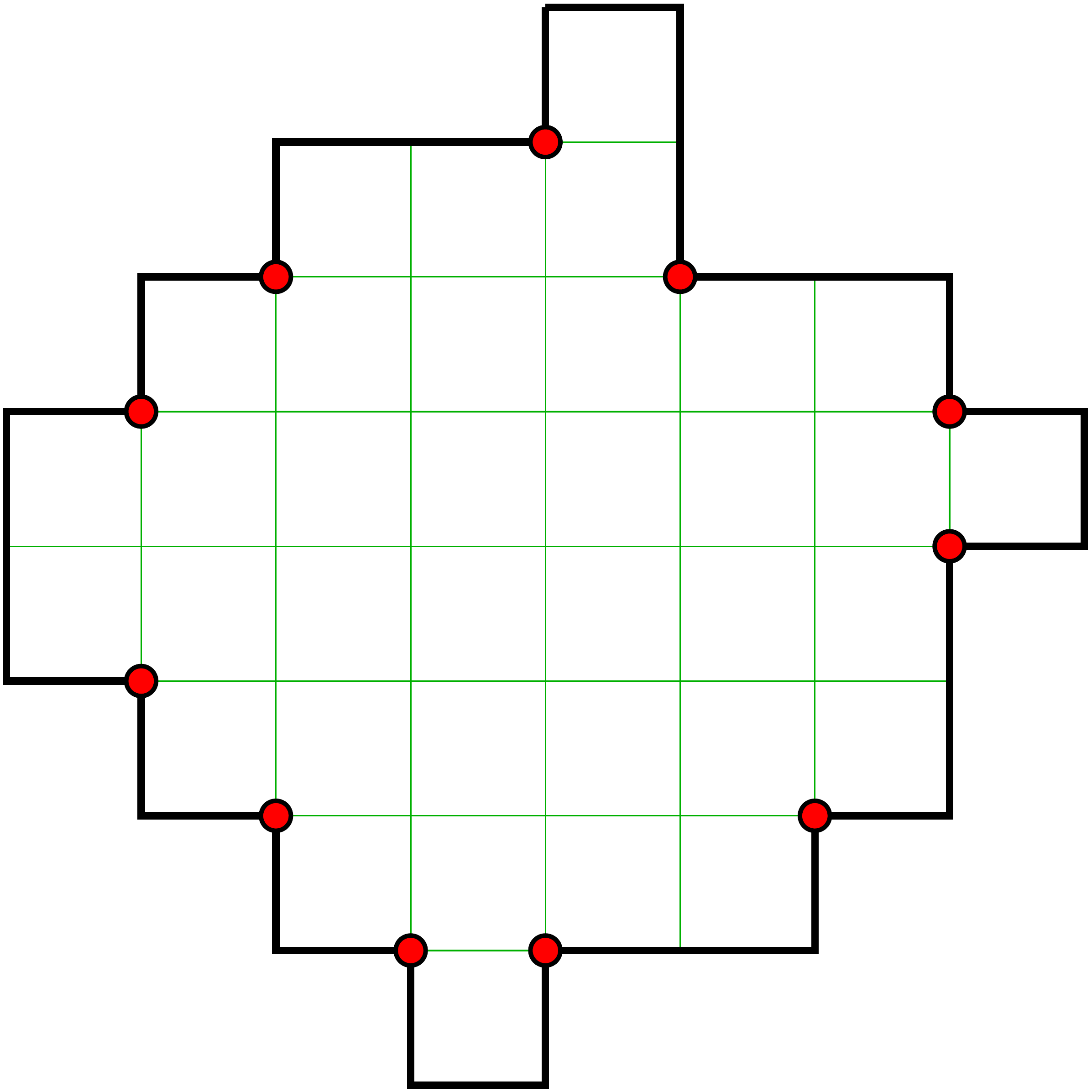}
         \caption{}
         \label{subfig:laminar_corner}
     \end{subfigure}
     \vskip2mm
     
     \begin{subfigure}[b]{0.19\textwidth}
         \centering
         \includegraphics[width=\textwidth]{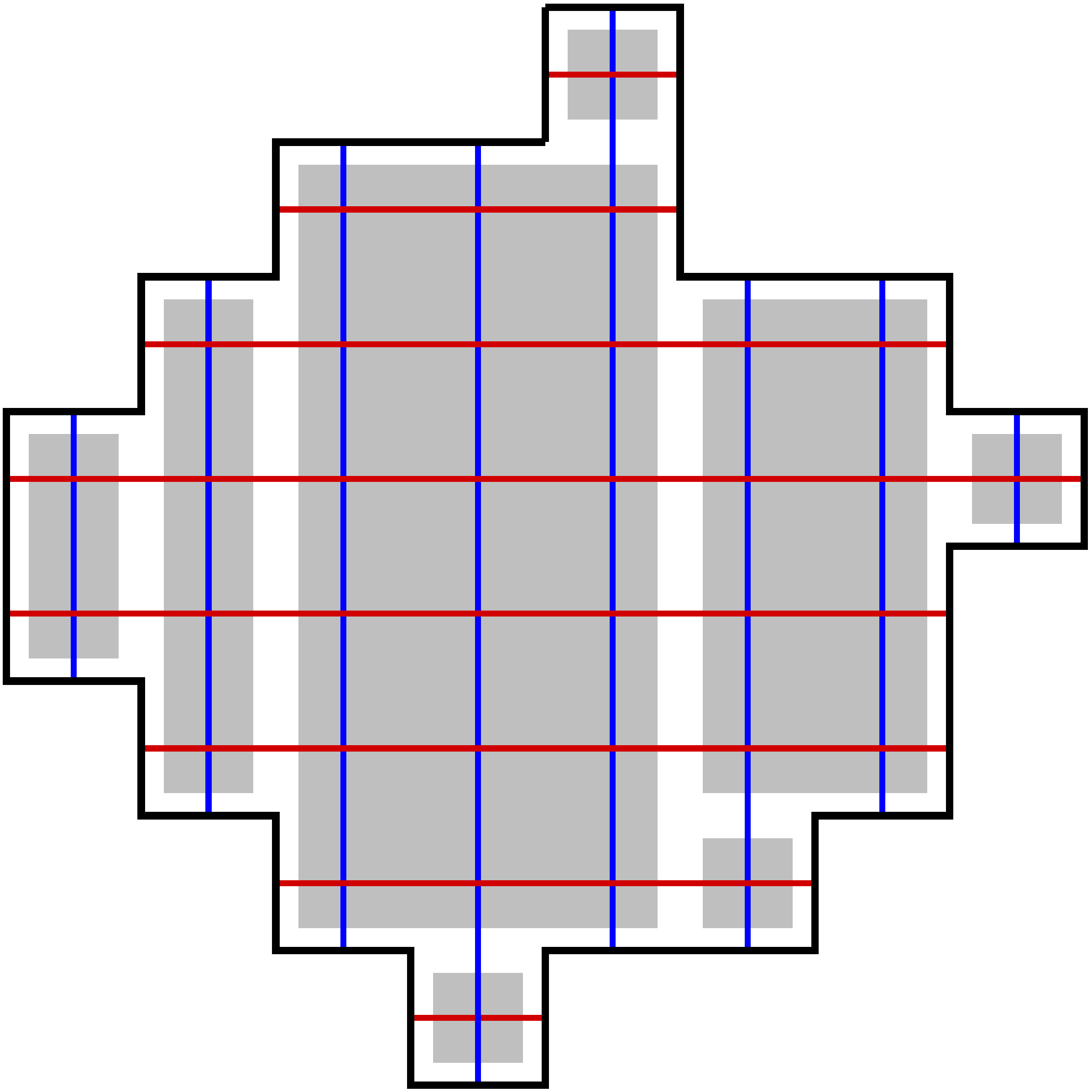}
         \caption{}
         \label{subfig:laminar_bundle}
     \end{subfigure}\hskip1cm
     \begin{subfigure}[b]{0.19\textwidth}
         \centering
         \includegraphics[width=\textwidth]{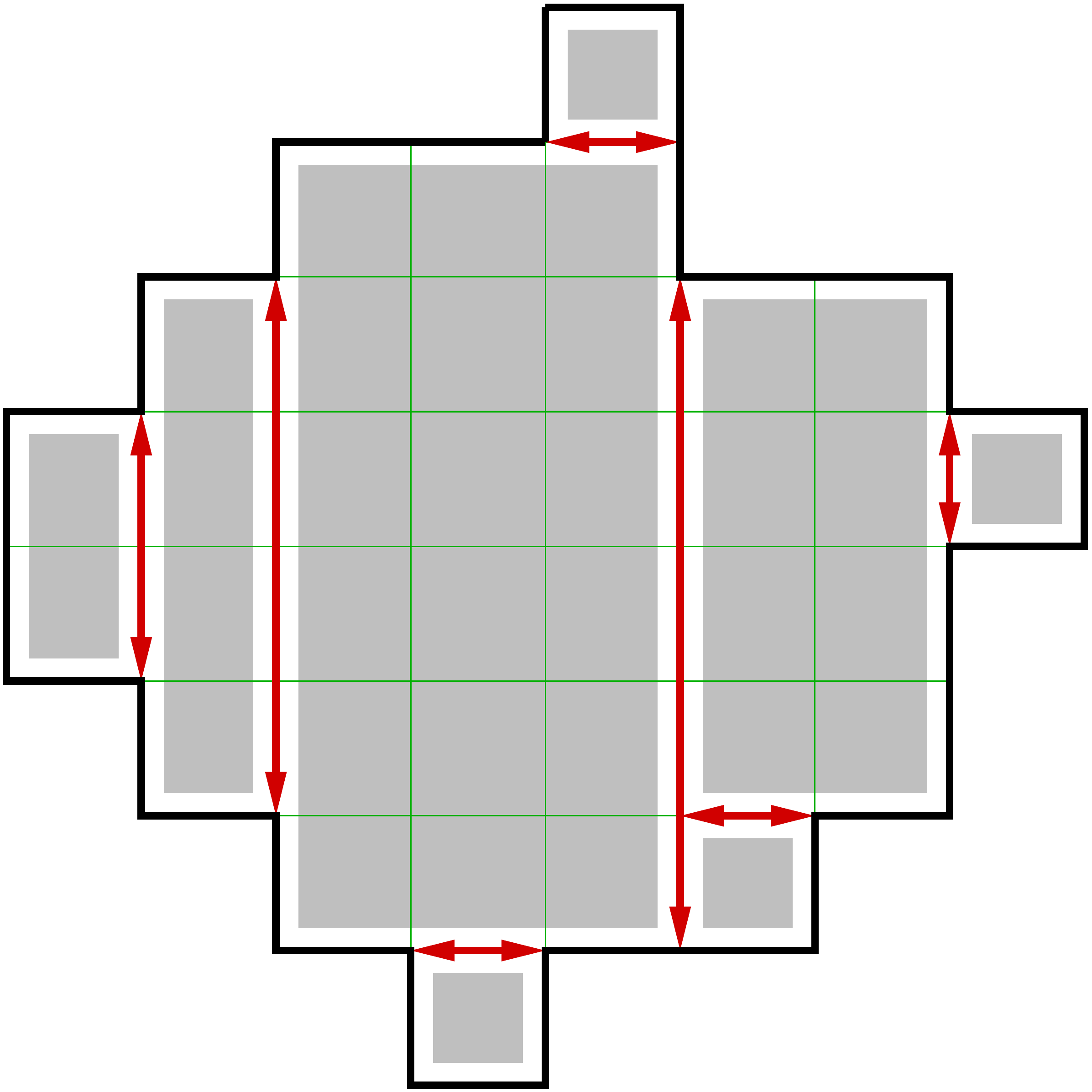}
         \caption{}
         \label{subfig:laminar_rectangle}
     \end{subfigure}\hskip1cm
     \begin{subfigure}[b]{0.19\textwidth}
         \centering
         \includegraphics[width=\textwidth]{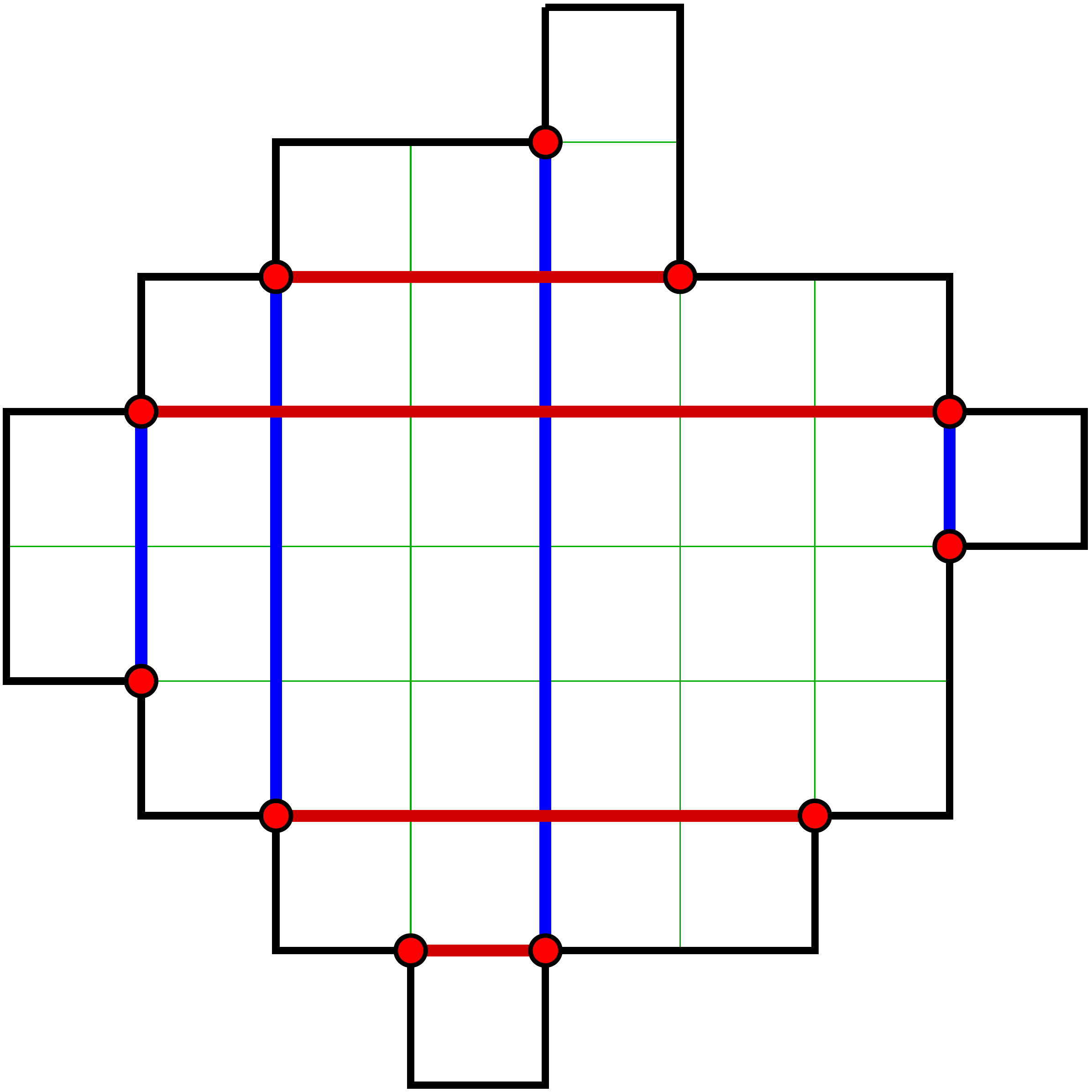}
         \caption{}
         \label{subfig:laminar_conflict}
     \end{subfigure}
\caption{Bundling of a bi-laminar instance.}
\end{figure}

A \term{laminar family} of chords is a collection of pairwise disjoint chords in a circle
such that for any three chords one of them separates the other two.
A \term{bi-laminar instance} is a circle together with ta red and a blue laminar family
of chords such that any two chords have at most one crossing (Figure \ref{subfig:laminar}).

A bi-laminar instance can be converted, by an appropriate
crossing-preserving transformation, into a family of blue vertical chords and
red horizontal chords drawn inside an \term{orthogonal polygon} $P$, i.e., a
polygon whose edges are parallel to the $x$- and $y$-axis (Figure
\ref{subfig:laminar_orth}).  Moreover, the polygon and the chords can be
chosen so that the chords are evenly spaced forming a regular grid inside $P$.

With the tranformed instance in hand, we can now subdivide the area inside $P$
into squares, each of them containing a crossing in its center. We will refer
to the green graph obtained from the union of $P$ and the perimeter of the
squares as the \term{dual net} of the instance
(Figure~\ref{subfig:laminar_corner}).  A bundled crossing corresponds to a
collection of squares in the dual net forming a \tem{rectangle} (Figures
\ref{subfig:laminar_bundle}-\ref{subfig:laminar_rectangle}); moreover, a
\tem{bundling} of the new instance corresponds to a partition of the squares
of the dual net into rectangles. Any such partition is called a
\term{rectangulation} of $P$, and a minimal bundling corresponds to a
rectangulation using a minimum number of rectangles.

The problem of rectangulating an orthogonal polygon into the minimum number of
rectangles was solved by at least three independent groups in the 80's,
Eppstein~\cite{Ep09} contains the relevant references. We next describe a
polynomial time algorithm for finding a minimal rectangulation.

The segments added to the polygon $P$ to obtain a rectangulation are either
horizontal or vertical (red bi-arrows in Figure
\ref{subfig:laminar_rectangle}). In a rectangulation with $R$ rectangles and
$S$ segments, the relation $R=S+1$ holds. Therefore, the problem of finding a
minimal rectangulation is equivalent to the problem of adding a minimal set of
segments inducing a rectangulation of~$P$.

Each concave corner $v$ of $P$ (red vertices in Figure
\ref{subfig:laminar_corner}) must be incident with at least one segment of any
rectangulation. We call this the \tem{requirement} at $v$ (the requirement is
related to what we later call the \tem{exponent} of $v$).  A \term{good
  segment} is a segment connecting two concave corners; it is called good because it
satisfies the requirement of two corners. Figure~\ref{subfig:laminar_conflict}
shows the given polygon together with all the good segments. The example shows
that pairs of segments may cross or share an endpoint, we say that they are in
\term{conflict}.

Ultimately the rectangulation problem for $P$ reduces to study a conflict
graph $C$ whose vertices are the possible good segments. Two good segments
in $C$ are joined by an edge, when they are in conflict. The conflict graph
is bipartite, with blue segments on one side and red segments on the
other. A minimum family of segments that
yields a rectangulation corresponds to a maximum independent set~$I$ of~$C$
plus a set of segments covering the corners which are not incident with
elements in $I$. Since computing a maximum independent set in a bipartite
graph can be done in polynomial time, the same holds true for computing a
minimum family of segments rectangulating $P$.
\medskip

Although we know how to find an exact solution for rectangulating $P$, we now
describe how to find an approximate solution to illustrate the algorithm we
will use to prove Theorem \ref{thm:main_8approx}.  Let $S$ be a set of
segments which is initially empty. Consider the concave corners of $P$ one by
one, when it comes to $v$ and $v$ is not incident to a segment $S$ choose a
direction $d$ (horizontal or vertical) and \emph{shoot a segment} in direction
$d$ from $v$, i.e., let $s_v$ be the segment with one end at $v$ extending
into the interior of $P$ and ending at the first point which belongs to a
segment in $S$ or the boundary of $P$, this segment $s_v$ is added to $S$.
From the discussion it should be clear that for laminar families of chords
this process yields a rectangulation that uses at most twice as many the
segments as an optimal one, i.e., it is a 2-approximation for the number of
segments and also for the number of rectangles. We refer to this approach as
the \term{greedy strategy}.  Fink et al.~\cite{FHSV16} analyze this strategy
in the setting where the input of the bundling problem is a circular
drawing. They show that it yields a~10-approximation for the number of
bundles. In Sections \ref{sec:approx_segments}-\ref{sec:approx_rectangles} we
analyze the greedy strategy for the bundling problem for good drawings and
show that this simple-minded strategy guarantees an $8$-approximation
(Theorem~\ref{thm:main_8approx}).

In Section \ref{sec:fivetwo_approx}, we study biparte collections of
strings; this is somehow closer to the laminar family studied here. There we
show that a slightly modified greedy strategy produces a solution which is a
$\frac{3}{2}$-approximation for the number of segments needed to rectangulate
$P$. The key is to first compute a set of segments of one color which
maximizes a parameter called marginal gain. This set is extended using the
greedy strategy until a rectangulation is obtained.


\section{Strings and Nets}

\begin{figure}[ht]
	\centering
	\begin{subfigure}[b]{0.46\textwidth}
         \centering
         \includegraphics[width=\textwidth]{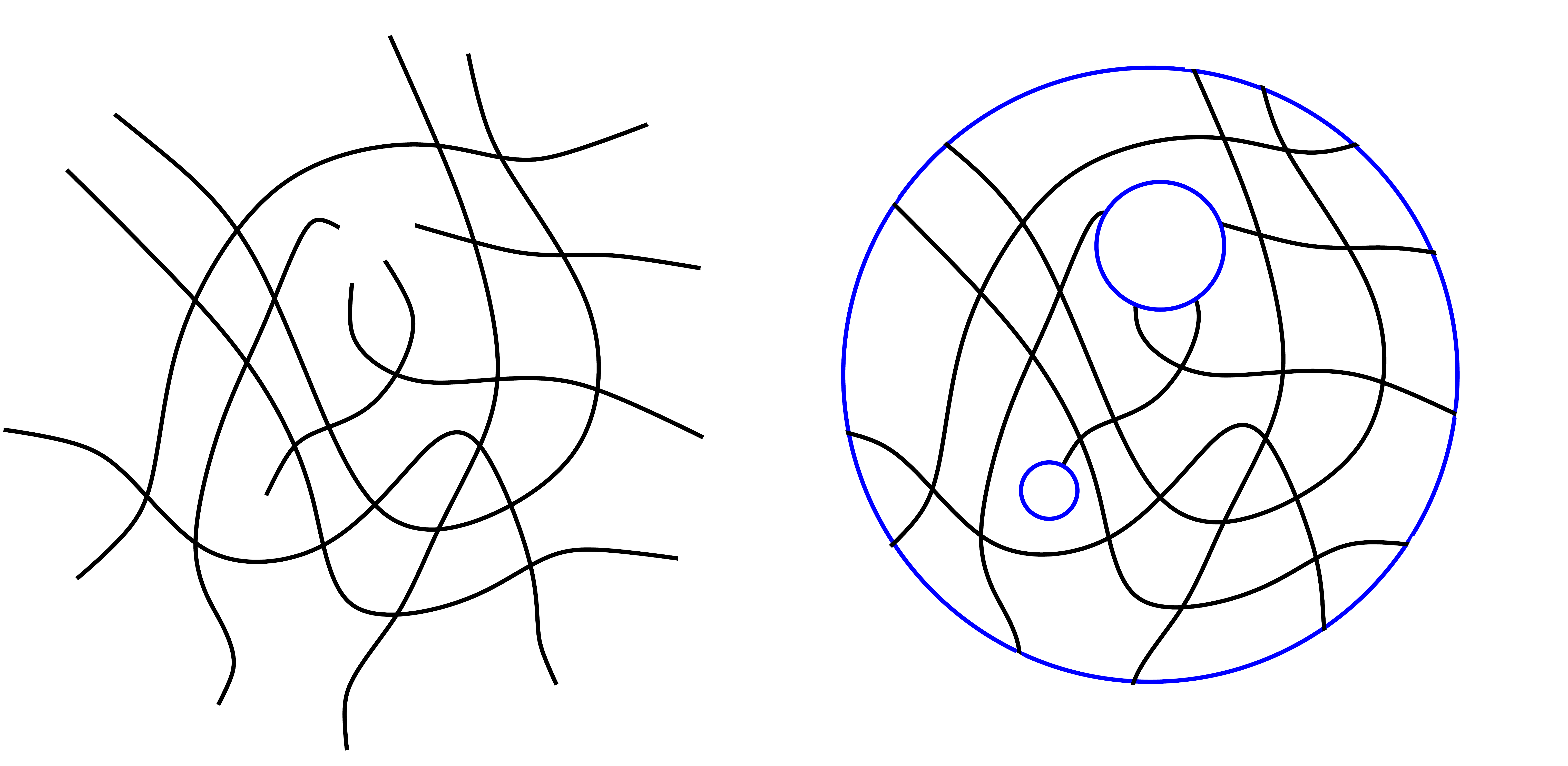}
         \caption{grounding strings}
         \label{subfig:grounding}
     \end{subfigure}
     ~
     \begin{subfigure}[b]{0.23\textwidth}
         \centering
         \includegraphics[width=\textwidth]{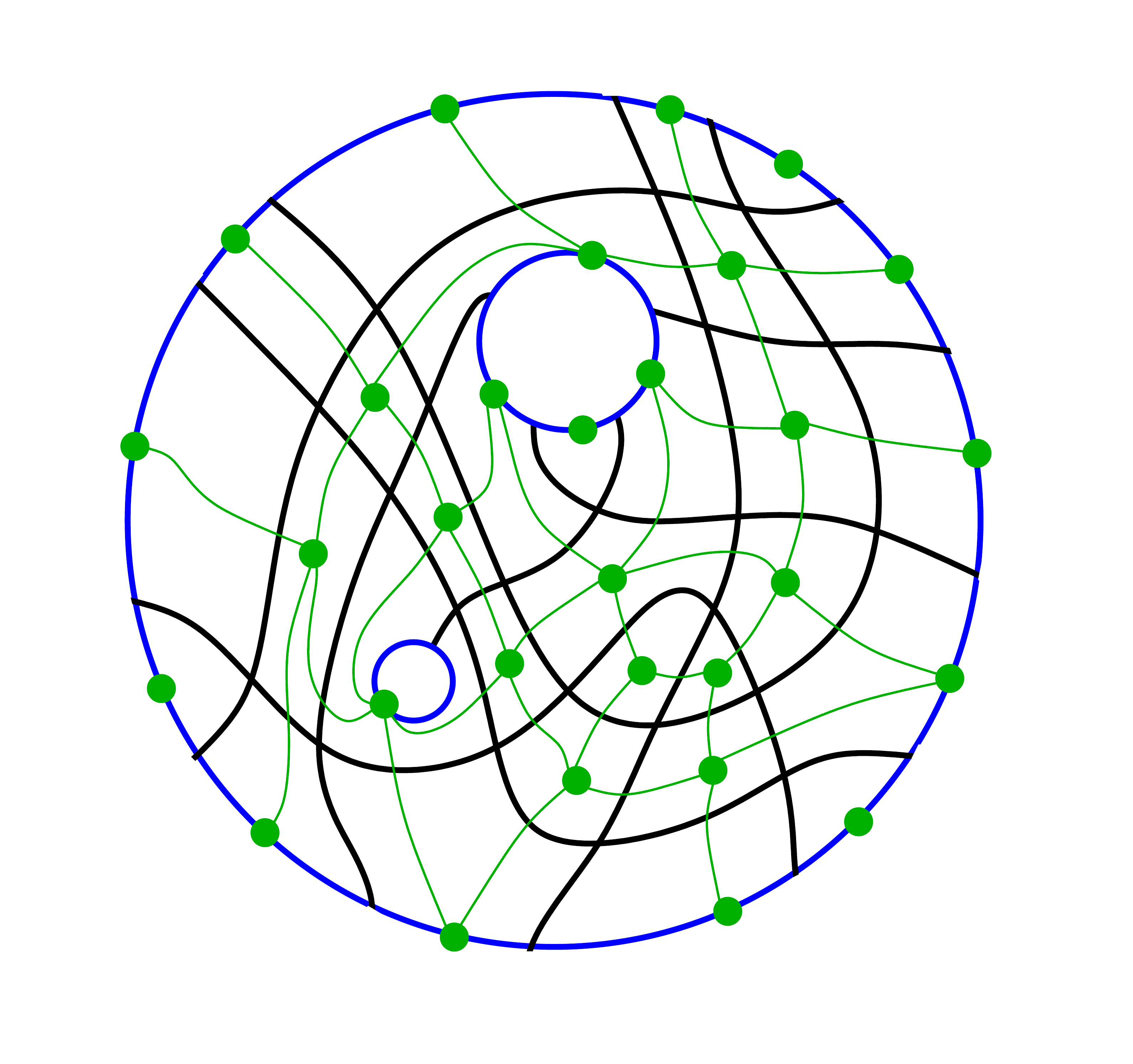}
         \caption{strings and net}
         \label{subfig:strings_and_net}
     \end{subfigure}
     ~
     \begin{subfigure}[b]{0.23\textwidth}
         \centering
         \includegraphics[width=\textwidth]{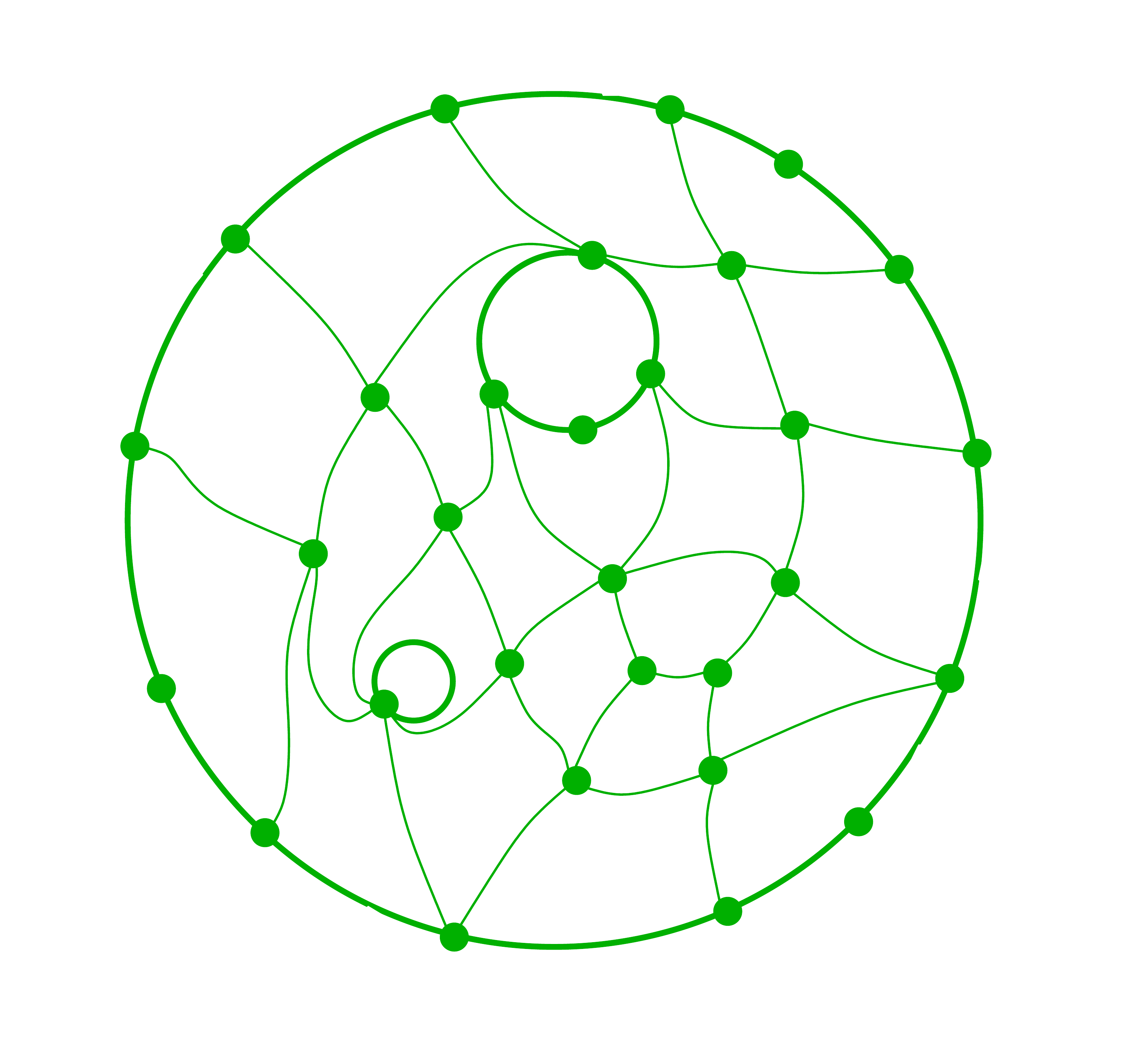}
         \caption{net}
         \label{subfig:net}
     \end{subfigure}
\caption{}
\end{figure}

We think of a bundling instance as a set of strings, i.e., as a set of simple
curves in the plane. Throughout this paper, unless otherwise stated, we assume
that the endpoints of any two strings are different and the union of the
strings forms a connected set.  A set $\EE$ of strings is \term{grounded} in
$\BB$ (see Figure~\ref{subfig:grounding}) if $\BB$ is a set of pairwise
disjoint blue simple closed-curves, we refer to them as \term{boundary
  curves}, such that (1) each string has its ends in the union of the boundary
curves; (2) boundary curves only intersect the strings at their ends; (3) each
boundary curve contains at least one end of a string; and (4) no two boundary
curves are incident with the same cell of $\EE\cup\BB$.

One can always turn a set of strings into a grounded one by adding a single
blue curve in each face where strings end (Figure \ref{subfig:grounding}).
Henceforth, unless otherwise stated, any set of strings we consider is
grounded.

To a grounded set of strings $(\EE,\BB)$ we associate a plane graph which
is called the \tem{dual net} or just the~\term{net}. The net $\mathcal{N}$ is
obtained by placing a vertex in each cell which is not bounded by a
single closed curve
in $\BB$ and by adding an edge between two vertices whenever
their corresponding cells share a segment of a string connecting two
consecutive intersection points.  We will draw our dual net as in
Figure~\ref{subfig:net} so that each vertex corresponding to a cell
incident to a boundary curve is drawn on the boundary curve, and edges
connecting two consecutive vertices in the same boundary curve are also
drawn along the boundary curve.

The faces of $\mathcal{N}$ come in two kinds: \term{boundary-holes}, defined as
the faces bounded by the boundary blue curves; and \tem{squares}, each of them
enclosing exactly one crossing of strings.

A bundle of a set of strings corresponds to a rectangle in the dual net. To be
more precise, a set of squares is a \tem{rectangle} if their enclosed
crossings induce a bundle, i.e., if the squares can be labeled with a
vertex-set of an $n\times m$-grid, so that any two squares adjacent in the
grid share an edge of their boundaries. A \term{rectangulation} is a partition
of the squares into rectangles, and such rectangulation is \tem{optimal} if it
corresponds to a bundling with a minimum number of bundles.

\section{Segments and Holes}
\label{sec:segments_holes}

In this section we consider a fixed set of grounded strings with its dual net
$\mathcal{N}=(V,E)$. The \term{border} of~$\mathcal{N}$ is the subgraph
induced by vertices and edges drawn on boundary curves. Any vertex or edge
of~$\mathcal{N}$ not on the border is \tem{interior}.  A degree-4 interior
vertex is called \tem{regular} and any other interior vertex is a
\tem{vertex-hole}. A \term{hole} is a vertex or face of $\mathcal{N}$ that
is either a vertex-hole or a boundary-hole.

A path of $\mathcal{N}$ is \tem{straight} if all its  non-end
vertices are regular and any two consecutive edges are opposite in the
rotation at their common vertex. A \term{cut-set} is a set $\mathcal{S}$ of edge-disjoint
straight-paths where every end of a path is either on a hole or in the
interior of another path of $\mathcal{S}$. We refer to the elements of a
cut-set as \term{segments}.

Given a rectangulation $\mathcal{R}$ of $\mathcal{N}$, an edge of $E$ is
\tem{separating} if it belongs to the boundary of two squares belonging to
different rectangles. A cut-set \tem{delimits} $\mathcal{R}$ if its segments
only include separating edges and each separating edge is included in a
segment. 

One can always iteratively build a cut-set delimiting $\mathcal{R}$: The
first segment $s_1$ is obtained by considering any edge $e_1$ in the set
$E'\subseteq E$ of separating edges, and by maximally extending $e_1$ into a
straight path.  In the $i$-th step, $s_i$ is obtained by maximally extending
an edge $e_i$ of $E'\setminus E(s_1\cup\cdots\cup s_{i-1})$ into a straight
path that is edge-disjoint from $s_1$,$\ldots$, $s_{i-1}$. This is done until
all the edges of $E'$ are covered by the segments.

If there is no regular vertex~$v$
whose four incident edges are separating, then the cut-set delimiting
$\mathcal{R}$ is unique. Otherwise, for any such vertex $v$, one can choose
the pair of opposite edges which belong to a common segment, the other
two segments end at $v$. This binary choice at any such vertex $v$ may give
rise to exponentially many delimiting cut-sets. In practice, we will not be
bothered by this technicallity. We choose and fix an arbitrary cut-set of
$\mathcal{R}$. With this in mind, we can now refer to the \term{segments of
  $\mathcal{R}$} as the segments of the chosen cut-set
delimiting~$\mathcal{R}$.

\begin{figure}[ht]
	\centering
     \begin{subfigure}[b]{0.33\textwidth}
         \centering
         \includegraphics[width=\textwidth]{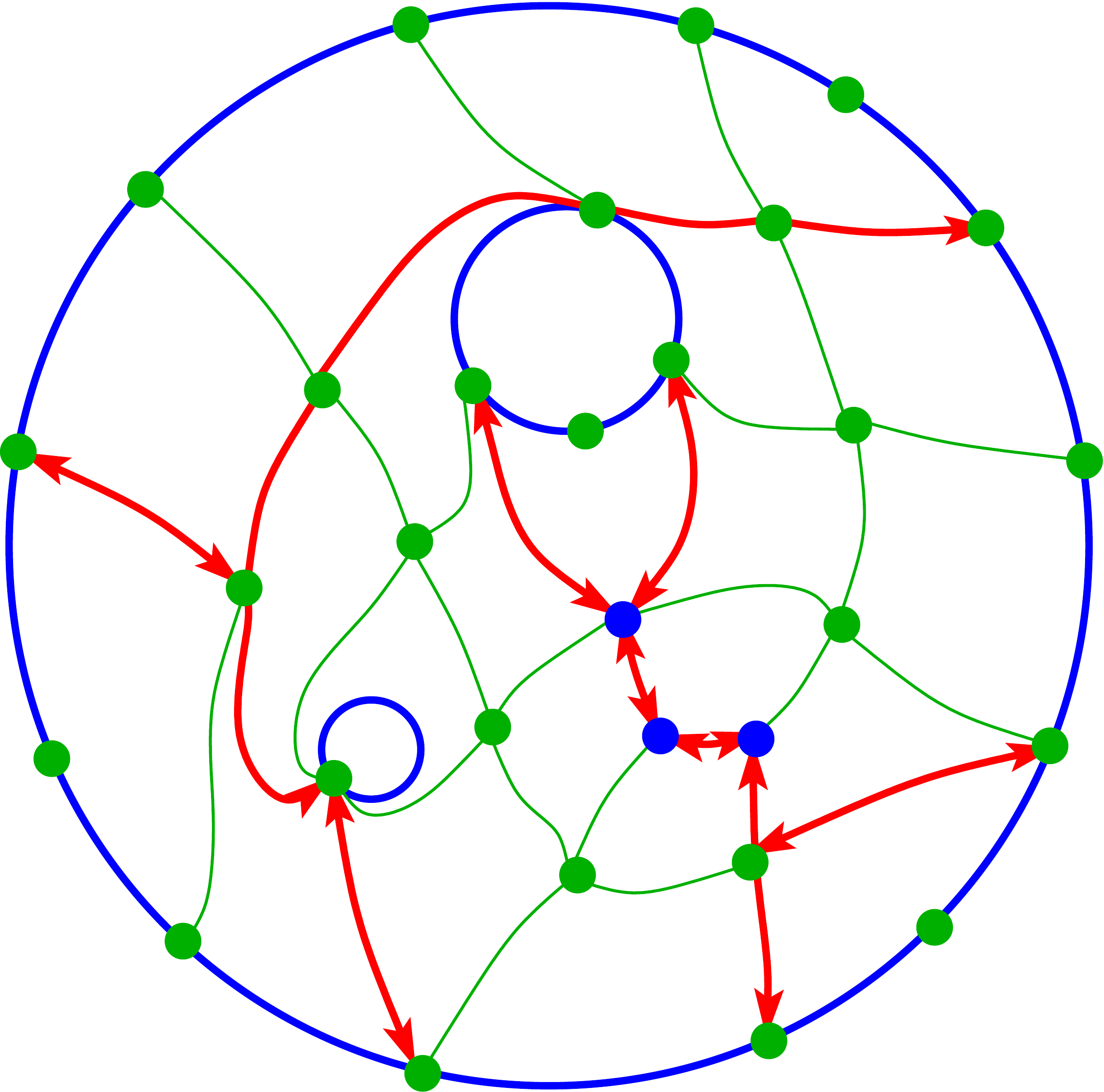}
         \caption{A rectangulation with 10 segments. Holes are blue.}
         \label{subfig:rect-net}
     \end{subfigure}
     \hskip15mm
     \begin{subfigure}[b]{0.35\textwidth}
         \centering
         \includegraphics[width=\textwidth]{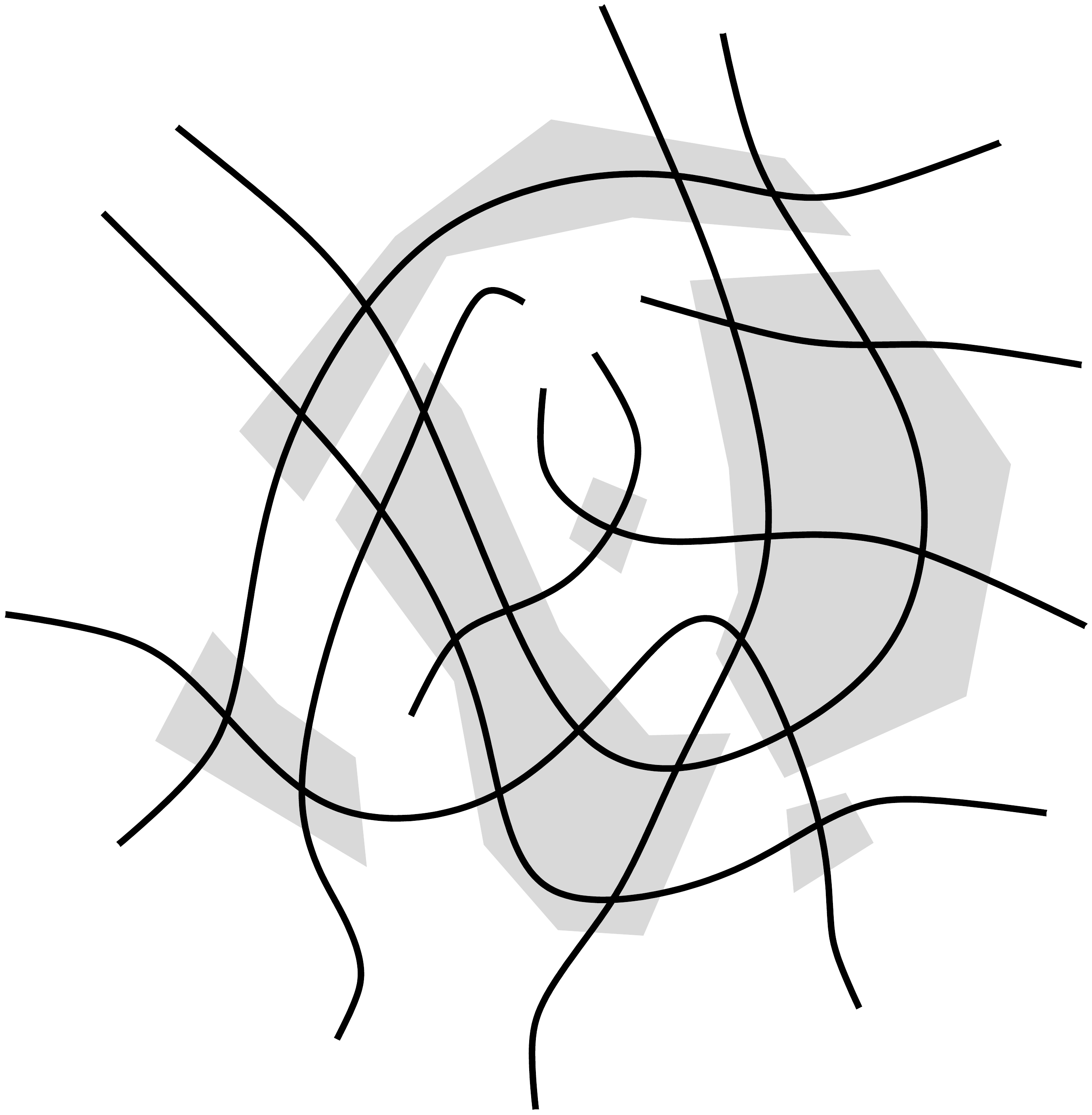}
         \caption{The corresponding bundling of the strings.}
         \label{subfig:bund-net}
     \end{subfigure}
\caption{}\vskip-3mm
\end{figure}

Figure~\ref{subfig:rect-net} shows a rectangulation with 10 segments and
Figure~\ref{subfig:bund-net} shows the corresponding bundling of
strings with 6 bundles.

Next we relate the number of rectangles and the number of segments in a
rectangulation.  Let $H=H(\mathcal{N})$ denote the number of holes of
$\mathcal{N}$.

\begin{lemma}\label{lemma:rectangles_segments_holes}
  In a net $\mathcal{N}$ with $H$ holes the numbers $R$ of rectangles and $S$
  of segments satisfy:
\begin{equation}\label{eq:rsh}
R-S+H=2\text{.} 
\end{equation} 
\end{lemma}

To prove Lemma \ref{lemma:rectangles_segments_holes}, to any
rectangulation~$\mathcal{R}$ of $\mathcal{N}$ we will associate a cubic plane
graph $\Gamma=\Gamma(\mathcal{R})$.  The construction will be used again in
Section~\ref{sec:R+H}.  \medskip

\noindent {\bfseries Construction of $\Gamma$:}
\label{constGamma}
First, we consider the subgraph $\mathcal{H}$ of $\mathcal{N}$ obtained from
the union of the segments in $\mathcal{R}$ and the border of
$\mathcal{N}$. Color the edges of $\mathcal{H}$ included in segments such that
for each segment all the edges on the segment have the same color and the
colors used for different segments are distinct.

\begin{figure}[ht]
  \centering
         \includegraphics[width=0.6\textwidth]{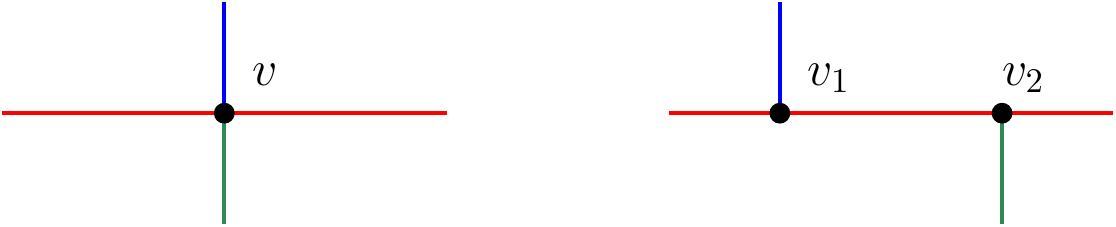}
\caption{Local transformation when $v$ is regular and $\text{deg}_{\mathcal{H}}(v)=4$.}
\label{subfig:transf_gamma_2}\vskip-3mm
\end{figure}

Next, we apply a local tranformation at each vertex $v\in V(\mathcal{H})$: If
$v$ is a regular vertex, then we keep $v$ unchanged unless
$\text{deg}_\mathcal{H}(v)=4$. In this case we split~$v$ into two vertices
$v_1,v_2$ joined by an edge as in Figure \ref{subfig:transf_gamma_2}, where
each $v_i$ has degree~$3$ and the color of $v_1v_2$ is the color of the unique
segment having $v$ in its interior.

If $v$ belongs to the border and $\text{deg}_\mathcal{H}(v)\geq 4$ or if $v$
is a vertex-hole of arbitrary degree we split $v$ as follows. Let
$\rho_\mathcal{H}(v)=(e_1,\ldots, e_d)$ be its rotation in $\mathcal{H}$, so
that when $v$ is in the border of $\mathcal{N}$, we assume that the edges
$e_1$ and $e_d$ are also in the border. Subdivide each vertex $e_i$ by adding
a degree-2 vertex in the middle of $e_i$. Remove $v$ and all the half-edges
incident with $v$. Next, add a path $(e_1,\ldots, e_d)$ or a cycle
$(e_1,\ldots, e_d, e_1)$ depending whether $v$ is in the border or not.

\begin{figure}[ht]
  \centering
   \includegraphics[width=0.3\textwidth]{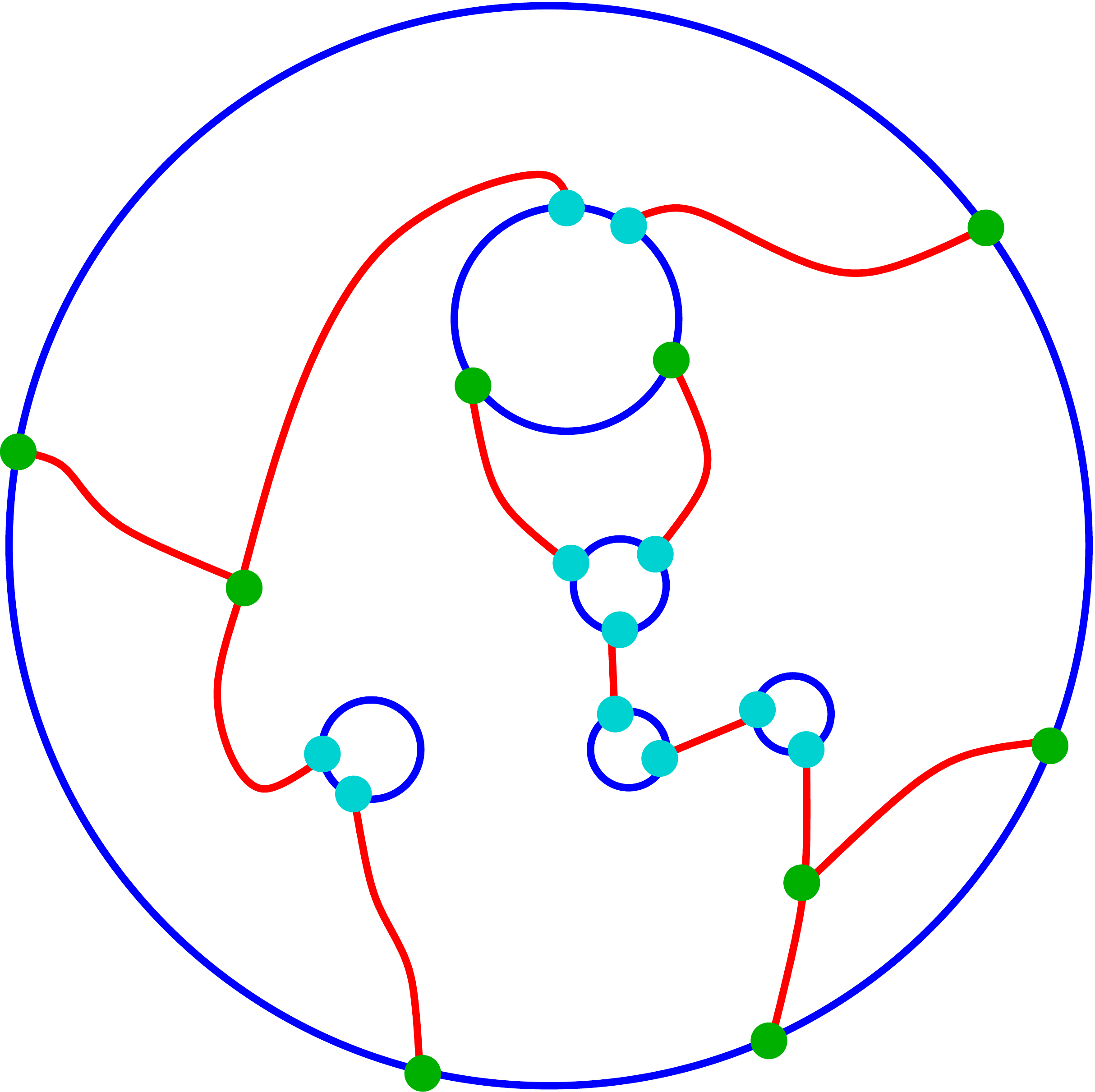}
   \caption{The graph $\Gamma$ corresponding to the rectangulation of
     Figure~\ref{subfig:rect-net}. Vertices obtained by splitting vertex-holes
     or vertices on boundaries are shown in light blue.}
   \label{fig:gamma}
\end{figure}

Finally, we  suppress degree-2 vertices, i.e., if $v$ is a vertex with
$\text{deg}_\mathcal{H}(v) = 2$ and incident edges $e_1,e_2$, then we
delete $v$ and make $e_1,e_2$ a single edge. This yields $\Gamma$.
Figure~\ref{fig:gamma} shows an example.

\begin{proof}[Proof of Lemma \ref{lemma:rectangles_segments_holes}]
  Given a rectangulation $\mathcal{R}$, let $\Gamma=\Gamma(\mathcal{R})$
  as above.  Each segment of the rectangulation corresponds to a
  monochromatic path in $\Gamma$ and each vertex of $\Gamma$ is an
  end-vertex of exactly one of them. Thus $|V(\Gamma)|=2S$. As $\Gamma$ is
  cubic, $E(\Gamma)=\frac{3}{2}|V(\Gamma)|=3S$. Finally, as each face of
  $\Gamma$ corresponds to either a rectangle or a hole, $\Gamma$ has $R+H$
  faces. Equation \ref{eq:rsh} now follows from Euler's formula.
\end{proof}

Before concluding this section, we make some remarks about $\Gamma$ that will
be used later. We let $\Gamma^*$ denote the plane dual of $\Gamma$.
\begin{remark}\label{rmk:gamma}
\begin{enumerate}[label=(\roman*)]
\item\label{it:faces_gamma}
  Each face of\/ $\Gamma$ corresponds to a hole or to a rectangle.
\item\label{it:independent_gamma}
  $\Gamma^*$ is a plane triangulation, i.e.,
  the boundary walk of each face consists of three edges.
\item\label{it:simple_gamma}
  $\Gamma$ and its planar dual\/ $\Gamma^*$ are
  simple graphs when every hole is incident with at least three segments of
  $\mathcal{R}$ (otherwise\/ $\Gamma$ has multiedges and/or loops).
\item\label{it:remark}
  The vertices of\/ $\Gamma^*$ corresponding to holes
  form an independent set in\/ $\Gamma^*$.
\end{enumerate}
\end{remark}

\section{Approximating the Number of Segments}  
\label{sec:approx_segments}

Let $\mathcal{N}=(V,E)$ be a net and $B\subset E$ be the set of edges in the
border of $\mathcal{N}$. An edge-set $E_0\subseteq E$ \term{saturates} a vertex
$v\in V$ if each angle induced by the edges of $E_0\cup B$ at $v$ sees at most
two squares, or, if $v$ is regular and no edge of $E_0$ is incident with
$v$. Moreover, $E_0$ \tem{saturates} $\mathcal{N}$ if every vertex is
saturated by $E_0$.
 
Naturally, the (edge-set of the) segments of a rectangulation saturate
$\mathcal{N}$.  The next lemma shows that saturation is also sufficient to
induce a rectangulation.

\begin{lemma}\label{lemma:saturating_is_rectangulating}
  A cut-set of a net $\mathcal{N}$ is saturating if and only if it delimits a
  rectangulation.
\end{lemma}

\begin{figure}[htb]
  \centering
     \begin{subfigure}[b]{0.44\textwidth}
         \centering
         \includegraphics[width=\textwidth]{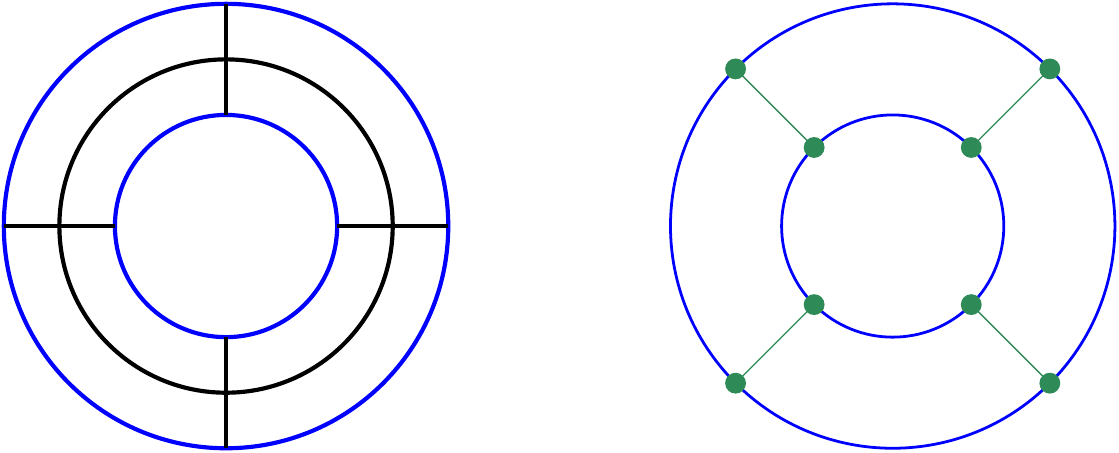}
         \caption{Square ring}
         \label{subfig:square_ring}
     \end{subfigure}
     \hskip12mm
     \begin{subfigure}[b]{0.44\textwidth}
         \centering
         \includegraphics[width=\textwidth]{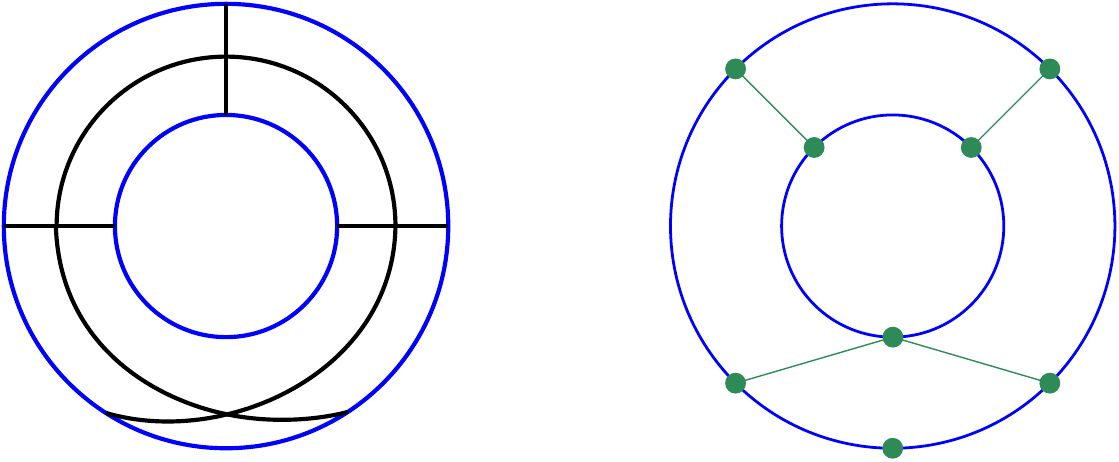}
         \caption{Square loop}
         \label{subfig:square_loop}{}
     \end{subfigure}
\caption{}\label{fig:ring_and_loop}\vskip-3mm
\end{figure}

Before proving Lemma \ref{lemma:saturating_is_rectangulating}, we observe that
this lemma does not extend to more general systems of curves. In Figure
\ref{fig:ring_and_loop} we depict more general systems of curves allowing
closed curves and self-intersecting strings. Their corresponding dual nets are
saturared by~$\emptyset$ whereas $\emptyset$ does not induce a rectangulation
in any of them.

The dual nets of Figure \ref{subfig:square_ring} and \ref{subfig:square_loop}
contain special configuration of squares forbidden in nets of strings: A
\tem{square-ring} is a circular sequence $(s_0,s_1,\ldots, s_m,s_0)$ of
squares, where $s_i$ is glued to $s_{i-1}$ and $s_{i+1}$ by using opposite
sides of $s_i$. A \tem{square-loop} is similar to a square-ring, except that
one square is glued by using two consecutive sides instead of opposite sides.

\begin{proof}[Proof of Lemma \ref{lemma:saturating_is_rectangulating}]
  If $\mathcal{S}$ is a set of segments delimiting a rectangulation, then each
  vertex is saturated by the edges of $\mathcal{S}$ whence the set of segments
  is saturating.

  Now let $\mathcal{S}$ be a saturating cut-set and let $B$ be the set of
  boundary-edges and $S$ be the set of edges covered by $\mathcal{S}$.  Also
  let $U\subset E\setminus(B\cup S)$ be the set of interior edges not covered
  by the segments in $\mathcal{S}$. We proceed by induction on $|U|$. If
  $|U|=0$, then each boundary edge of each square is either in the border or
  contained in a segment of $\mathcal{S}$. In this case $\mathcal{S}$ induces
  a rectangulation.

  Now let $F$ be a face of the subgraph induced by $B \cup S$, such that~$F$
  contains at least one edge from $U$.  We have to show that $F$ is a
  rectangle. Let $(e_1,\ldots, e_m)$ be the counterclockwise boundary walk of
  $F$ (where $e_i$ is the $i$-th edge encountered in the walk).  Let $e_0=e_m$
  and define $W=(e_0,e_1,\ldots, e_m)$.  Since $\SS$ is saturating, at most
  one edge of $U$ is pointing into the interior of $F$ between any pair $e_i$
  and $e_{i+1}$ of edges. A square of $\NN$ is a \term{corner} of $F$, if it
  is contained in $F$ and two of its boundary edges are consecutive in $W$.
  Since the squares incident to the edges in $W$ do not form a square-ring not
  a square-loop, it follows that $F$ has at least two corner squares.

  Let $s$ and $s'$ be consecutive corners along $W$ and let $u_1,\ldots,u_r$
  be the sequence of edges pointing into~$F$ between $s$ and $s'$ and let
  $u_0$ and $u_{r+1}$ be the edges opposite to $u_1$ and $u_r$ on $s$ and $s'$
  respectively. Note that $u_0$ and $u_{r+1}$ belong to $W$. Let $s_0=s$ and
  $s_r=s'$ and for $1\leq i < r$ let $s_i$ be the square incident to $u_i$ and
  $u_{i+1}$. By relabeling $W$, we may assume that edge $e_i$ is incident with
  $s_i$ for $i=0,\ldots, r$.  We let $e_i'$ be the edge opposite to $e_i$ in
  $s_i$. If an $e_i'$ edge is in $U$, then all the $e_i'$ edges must be in
  $U$, or else a vertex incident with two $e_i'$s, one in $U$ and one not in
  $U$, would not be saturated by $\mathcal{S}$.

  If all the $e_i'$ edges are in $B\cup S$, then
  $W = (e_1,\ldots, e_r,u_{r+1},e_r',\ldots, e_1',u_0)$ and
  $F = \bigcup_{i=0}^r s_i$. In this case $F$ is a rectangle.  Otherwise all
  the $e_i'$ edges are in $U$. In this case we add to $\mathcal{S}$ the
  segment obtained from the union of the~$e_i'$ edges to obtain $\SS'$. The
  ends of this new segment have already been saturated by $\mathcal{S}$ and
  all its interior vertices are saturated by the segment, hence $\SS'$ is
  saturating. By induction~$\SS'$ delimits a rectangulation.  One rectangle of
  this rectangulation is $F = \bigcup_{i=0}^r s_i$. Let $F'$ be the rectangle
  covering the square on the other side of $e_1'$. For each $i$ let $s_i'$ be
  the square oposite of $s_i$ at $e_i'$. The saturation property of the
  vertices on the path $e_1',\ldots,e_r'$ implies that all the interior
  vertices of this path are of degree four in the net with all incident edges
  in $U$. At the end vertices of the path we find that~$s_1'$ and $s_r'$ are
  consecutive corner squares of $F'$. Hence the union of $F$ and $F'$ is a
  rectangle delimited by~$\mathcal{S}$.
\end{proof}
\enlargethispage{3.8mm}

\begin{definition}[Exponent]
  Given a net $\mathcal{N}=(V,E)$ and $v\in V$, the exponent of $v$ is the
  minimum number of edges in an edge-set saturating $v$. Hence
  $\text{exp}(v) = 0$ if $v$ is a regular vertex and 

 \[ \text{exp}(v)=\begin{cases} 
     \rule[-5mm]{0mm}{5mm}
     \left\lfloor \frac{\text{deg}_\mathcal{N}(v)}{2} \right\rfloor-1
                  & v \text{ is in the border;}\\
     \rule[-5mm]{0mm}{5mm}
     \left\lceil \frac{\text{deg}_\mathcal{N}(v)}{2} \right\rceil
     &  v \text{ is a vertex-hole.}
   \end{cases}
\]
We let  $\text{exp}(\mathcal{N}):=\sum_{v\in V}\text{exp}(v)$.  
\end{definition}
\goodbreak

\noindent {\bfseries A greedy strategy:}
This strategy consists on linearly ordering the vertices $v_1,\ldots, v_k$
of~$\NN$ with positive exponent and start adding segments at the vertices with
increasing index.  When it comes to $v_i$ some incident edges may already be contained
in segments belonging to earlier vertices. Select a minimal set of edges
not covered by the segments such that shooting segments from these edges
results in an edge set saturating $v_i$. Note that the number of segments
introduced to saturate $v_i$ is upper bounded by~$\text{exp}(v_i)$.
\medskip

Henceforth, we will denote the number of rectangles and segments in an optimal
rectangulation of~$\mathcal{N}$ by $R_{opt}$ and $S_{opt}$,
respectively. Likewise, we let $R_{greed}$ and $S_{greed}$ be the number of
rectangles and segments, respectively, obtained after a run of the greedy
strategy in $\mathcal{N}$ for some linear order of the vertices.

\begin{observation}\label{obs:approx_seg}
The following hold true for a net $\mathcal{N}$:
\begin{enumerate}[label=(\roman*)]
	\item \label{it:seg_greed} $S_{greed}\leq \text{exp}(\mathcal{N})$;
	\item \label{it:seg_opt} $S_{opt}\geq  \frac{1}{2}\text{exp}(\mathcal{N})$; and consequently
	\item  \label{it:seg_greed_vs_opt}$S_{greed}\leq 2\cdot S_{opt}$.
\end{enumerate}
\end{observation}
\begin{proof}
  Item \ref{it:seg_greed} directly follows from the definition of the greedy
  strategy. Since each segment of a rectangulation contains at most two
  vertices of positive exponent, \ref{it:seg_opt} holds because a segment
  has only two ends.
\end{proof}

\section{Rectangles and Holes}
\label{sec:R+H}

In this section we collect a few facts about rectangles and holes that will be
used in our approximations. It is important to observe that a greedy
rectangulation approximates the optimal when $R_{greed}$ is bounded by a
constact factor of $R_{opt}$.  The next observation already gives a related
bound by adding the holes.

\begin{observation}\label{obs:greed_bound_by_rec_holes}
  $R_{greed}\leq 2R_{opt}+H-2$.
\end{observation}
\begin{proof}
  Apply Lemma~\ref{lemma:rectangles_segments_holes} to both sides of
  Observation \ref{obs:approx_seg}.\ref{it:seg_greed_vs_opt}.
\end{proof}

Our task of approximating $R_{opt}$ now reduces to understand under which
circumstances $H=O(R_{opt})$. Let us start by deriving a bound for the odd
holes. Let $H_{odd}$ be the number of vertex-holes with odd degree in
$\mathcal{N}$, i.e., $H_{odd}$ is the number of vertices of odd degree
$\geq 5$ in $\mathcal{N}$.

\begin{observation} \label{obs:bound_odd_holes}
 $H_{odd}\leq 4\cdot R_{opt}$.
\end{observation}

\begin{proof}
  Consider an optimal rectangulation with a set of segments. If $v$ is a
  vertex-hole of odd-degree, then at least one angle induced by the segments
  at $v$ sees exactly one square. This square is the corner of a
  rectangle. Since each rectangle has $4$ corners,
  $H_{odd}\leq 4\cdot R_{opt}$.
\end{proof}

Our next observation about holes requires the following general
observation about triangulations:

\begin{observation}\label{obs:independent_triangulation}
  In a simple plane triangulation with $n$ vertices, an independent set has
  size at most $\frac{2}{3}n$.
\end{observation}
\begin{proof}
  The degree of a vertex in a simple triangulation is at least 3.  Hence, each
  vertex of the independent set is incident to at least three triangles. Each
  triangle sees at most one vertex from the independent set $I$. Hence we get
  $3|I| \leq |F| = 2(n-2)$, whence $|I| \leq \frac{2}{3}n$.
\end{proof}

\begin{definition}
  Given a rectangulation $\mathcal{R}$ of $\mathcal{N}$,
  $\delta(\mathcal{R})$ denotes the minimum number of segments of
  $\mathcal{R}$ incident to any hole of $\mathcal{H}$ ($\delta(\mathcal{R})$
  is the same as the minimum degree among the vertices of $\Gamma^*(\mathcal{R})$ that represent holes).
   We let
  $\delta(\mathcal{N})$ be the minimum integer $k$ for which there is an
  optimal rectangulation $\mathcal{R}_0$ of $\mathcal{N}$ with
  $\delta(\mathcal{R}_0)=k$.
\end{definition}

\begin{figure}[ht]
    \centering
     \begin{subfigure}[b]{0.2\textwidth}
         \centering
         \includegraphics[width=\textwidth]{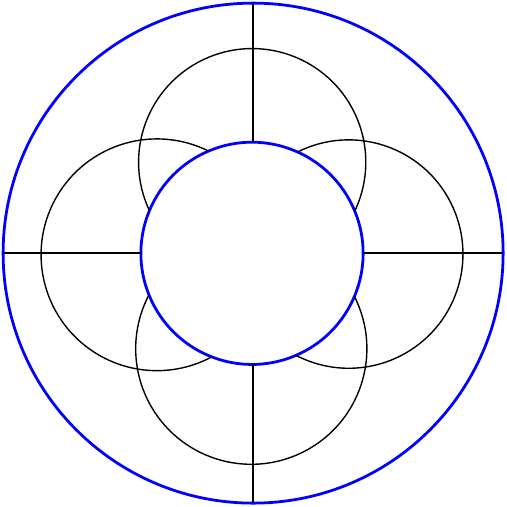}
         \caption{}
         \label{subfig:example_delta1}
     \end{subfigure}
     \hskip15mm
     \begin{subfigure}[b]{0.2\textwidth}
         \centering
         \includegraphics[width=\textwidth]{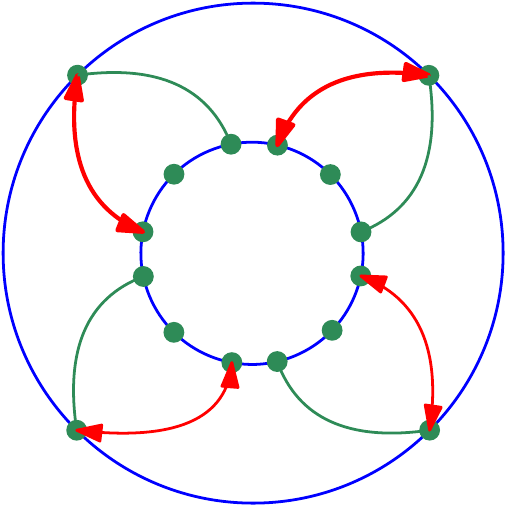}
         \caption{}
         \label{subfig:example_delta2}
     \end{subfigure}
     \caption{A set of strings (a) and its dual net (b).
       The red segments induce an optimal rectangulation, indeed $\delta(\mathcal{N})=4$.
       The exponents only imply $\delta(\mathcal{N})\geq 0$.}\vskip-3mm
\label{fig:example_delta}
\end{figure}

Note that when $h$ is a vertex-hole, then
$\deg_{\Gamma^*(\mathcal{R})}(h)\geq \text{exp}(h)$. For boundary-holes a
corresponding lower bound is given by the sum of exponents of the incident
vertices. Figure~\ref{fig:example_delta} shows that there are examples where
$\delta(\mathcal{N})$ is much larger than given by these lower bounds.

\begin{observation}\label{obs:delta_geq_3}
  If $\delta(\mathcal{N})\geq 3$, then $H\leq 2\cdot R_{opt}$.
\end{observation}
\begin{proof}
  Let $\mathcal{R}_{opt}$ be an optimal rectangulation with $\delta(\mathcal{R})=\delta(\mathcal{N})$ and let
  $\Gamma=\Gamma(\mathcal{R})$ be the 3 regular helper graph introduced in
  Section \ref{sec:segments_holes}, page~\pageref{constGamma}.  Condition
  $\delta(\mathcal{N})\geq 3$ and Remark \ref{rmk:gamma} imply that the holes
  of $\mathcal{N}$ correspond to vertices of an independent set of the simple
  plane triangulation $\Gamma^*$.  Observation
  \ref{obs:independent_triangulation} and Remark
  \ref{rmk:gamma}.\ref{it:faces_gamma} imply
  $H\leq \frac{2}{3}|V(\Gamma^*)|= \frac{2}{3}(H+R_{opt})$. Consequently
  $H\leq 2\cdot R_{opt}$.
\end{proof}

\subsection{Toothed-holes and toothed-faces}

 \begin{figure}[th]
    \centering
    \includegraphics[width=0.4\textwidth]{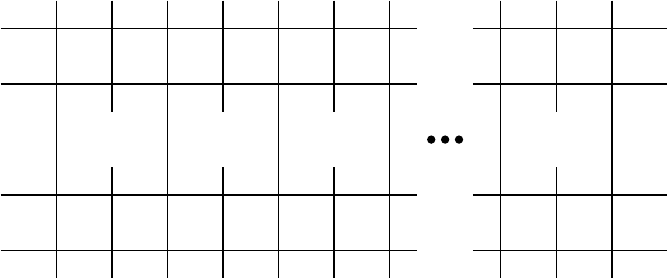}
         \caption{An example where $R_{opt}=2$ and $H$ is arbitrarily large.}
         \label{fig:many_tooth_faces}
     \end{figure}

     Previously, we saw that if all holes are incident with at least three
     segments, then $H$ is $O(R_{opt})$. Unfortunately, it is not true in
     general that $H$ is $O(R_{opt})$: Figure \ref{fig:many_tooth_faces} shows
     that $H$ can be arbitrarily large compared to $R_{opt}$. With the next lemma
     we prove that the unboundedness of $H$ in terms of $R_{opt}$ can
     always be attributed to the presence of toothed-faces.

\begin{lemma}\label{lemma:holes_bounded_by_t-faces}
$H\leq 6R_{opt}+t(\EE)-4$.
\end{lemma}

\begin{figure}[ht]
  \centering
     \begin{subfigure}[b]{0.4\textwidth}
         \centering
         \includegraphics[width=\textwidth]{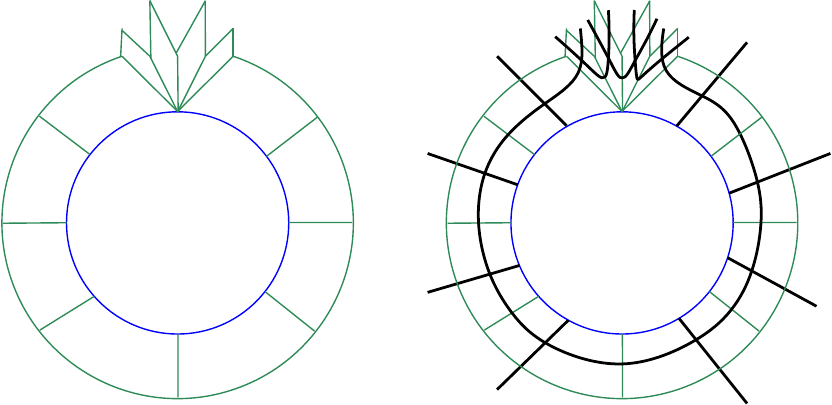}
         \caption{}
         \label{subfig:singular1}
     \end{subfigure}
     \hskip15mm
     \begin{subfigure}[b]{0.4\textwidth}
         \centering
         \includegraphics[width=\textwidth]{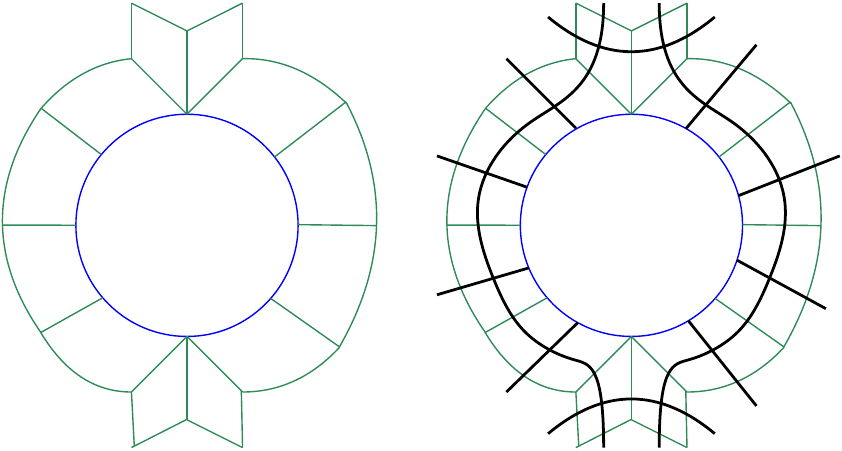}
         \caption{}
         \label{subfig:singular2}
     \end{subfigure}

     \caption{The two kinds of toothed-holes}
     \label{fig:singular_holes}

\end{figure}

As we are now dealing with nets, let us introduce the dual counterpart of a 
toothed-face:

\begin{definition} A \term{toothed-hole} of $\mathcal{N}$ is a boundary-hole
  $h$ for which all the vertices on its boundary satisfy one of the two
  conditions:(a) one vertex has degree 7 while the rest has degree 3; or (b)
  two vertices have degree 5 while the rest has degree 3. See Figure
  \ref{fig:singular_holes}.  We let $t(\mathcal{N})$ the number of
  toothed-holes in $\mathcal{N}$ (equal to $t(\EE)$ the number of
  toothed-faces in $\EE$).
\end{definition}

Given a rectangulation $\mathcal{R}$, a hole $h$ is {\em incident to a corner} if there is a vertex $v$ in the net 
which is a corner of a rectangle of $\mathcal{R}$ and either $h=v$ (vertex-hole) of $v$ is a vertex on the boundary of $h$
(boundary-hole).



\begin{lemma} \label{lm:bound_holes_pseudosegments}
Let $\mathcal{N}$ be the dual net of a set of pseudosegments with an optimal rectangulation $\mathcal{R}_{opt}$. Suppose that $H_2$ is the number of holes incident with at most two segments. Then the following hold:

\begin{itemize}
\item[(i)] $H_2\leq 4\cdot R_{opt}+t(\mathcal{N})$; and
\item[(ii)] if $R_{opt}\geq 2$ then every hole is incident with at least two segments of $\mathcal{R}_{opt}$.
\end{itemize} 
\end{lemma}


\begin{figure}[ht]
  \centering
     \begin{subfigure}[b]{0.2\textwidth}
         \centering
         \includegraphics[width=\textwidth]{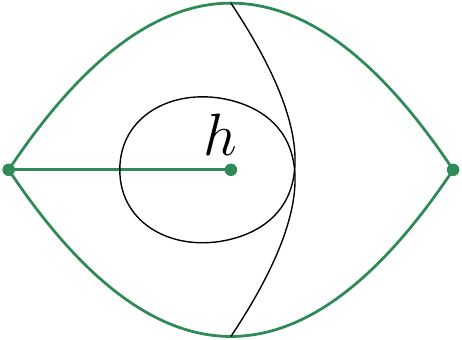}
         \caption{}
         \label{subfig:vertex_hole_low_deg1}
     \end{subfigure}
     \hskip15mm
     \begin{subfigure}[b]{0.2\textwidth}
         \centering
         \includegraphics[width=\textwidth]{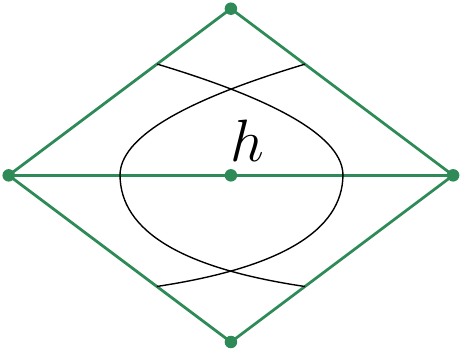}
         \caption{}
         \label{subfig:vertex_hole_low_deg2}{}
     \end{subfigure}
     \hskip15mm
          \begin{subfigure}[b]{0.15\textwidth}
         \centering
         \includegraphics[width=\textwidth]{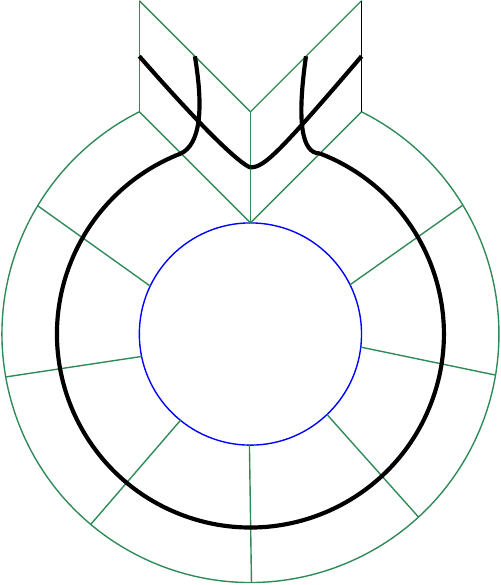}
         \caption{}
         \label{subfig:forb_boundary_hole}{}
     \end{subfigure}

     \caption{Three forbidden situations in the dual net of an arrangement of pseudosegments}
   \end{figure}
   
\begin{proof}

  Let $h$ be a hole incident with at most two segments but not incident with a
  corner. We will show that $h$ is toothed. This is enough to prove (i)
  because then every hole counted in $H_2$ is either incident to a corner (and
  there are at most $4R_{opt}$ of them) or is toothed.

  First, suppose by contradiction that $h$ is a vertex-hole.  How small can
  $\text{deg}_{\mathcal{N}}(h)$ be?  If $\text{deg}_{\mathcal{N}}(h)=1$, then
  in the only square incident to $v$ contains a self-intersecting string as
  shown in Fig.\ref{subfig:vertex_hole_low_deg1}, such a string is not a
  pseudosegment. If $\text{deg}_{\mathcal{N}}(h)=2$, then two squares incident
  with $v$ contain two pieces of strings forming a lense as shown in Figure
  \ref{subfig:vertex_hole_low_deg2}, this is not allowed for pseudosegments.
  If $\text{deg}_{\mathcal{N}}(h)\geq 3$, then $\text{deg}_{\mathcal{N}}(h)$
  cannot be odd, as otherwise $h$ would be incident to a corner.  Hence
  $\text{deg}_{\mathcal{N}}(h)$ must be even, and because $h$ is a hole,
  $\text{deg}_{\mathcal{N}}(h)\geq 6$. However, this implies that
  $\text{exp}(h)\geq 3$, showing that $h$ is incident with at least three
  segments, a contradiction. Thus $h$ is a boundary hole.

  Consider the vertices of $\mathcal{N}$ on the boundary $\partial h$ of
  $h$. Since $h$ is incident to no corner and to at most $2$ segments, the
  vertices on $\partial h$ have odd degree and the sum of their exponents is
  at most $2$. This restricts the degrees of the vertices on $\partial h$ to
  be arranged as one of the following kinds: (a)~exacly one vertex has degree
  7 while the rest has degree 3; (b)~two vertices have degree 5 while the rest
  has degree 3; (c)~exacly one vertex has degree $5$ while the rest has degree
  $3$; (d) all vertices have degree $3$. If either (a) and (b) occurs, then
  $h$ is toothed; it remains to show that neither (c) nor (d) occurs.

  If (d) occurs, then the squares incident with $h$ would form a square-ring,
  which would correspond to a cyclically closed string, not a pseudosegment .
  If (c) occurs, then, as illustrated in
  Figure~\ref{subfig:forb_boundary_hole}, the squares incident to the vertices
  in $h$ would induce a pair of strings that cross at least twice, a
  contradiction. Thus, $h$ is toothed, ultimately implying (i).

  Now we turn into proving (ii). For contradiction, suppose that
  $R_{opt}\geq 2$ and that $\mathcal{N}$ has a hole~$h$ incident with less
  than $2$ segments. As shown in (i), it follows that $h$ must be a
  boundary-hole. Since~$h$ is incident with less than $2$ segments there is a
  rectangle $r$ in the rectangulation whose boundary contains~$\partial h$.

  If $h$ has no incident segments, then $\partial h$ equals the boundary of
  $r$ whence all the pseudosegments ending on $h$ form a single bundle. From
  the connectivity assumption for the set of pseudosegments it then follows
  that $R_{opt}=1$, a contradiction.

  Now suppose that $h$ has a single incident segment $s$. There is a rectangle
  $r$ in the rectangulation whose boundary contains $\partial h$ and continues
  along $s$ on both sides.  Let $\sigma\in \EE$ be the first string
  crossing~$s$ when~$s$ is oriented away from its end in $h$. The intersection
  of $\sigma$ with the rectangle $r$ can only consist of one connected piece,
  otherwise $\sigma$ would have a selfintersection in $r$ or have two
  crossings with some other pseudosegment. This, however, implies that
  $\sigma$ is a closed loop and not a pseudosegment.  Hence every hole is
  incident to at least two segments.
\end{proof}

\begin{proof}[Proof of Lemma \ref{lemma:holes_bounded_by_t-faces}]
  If $R_{opt}=1$, then Lemma \ref{lemma:holes_bounded_by_t-faces} holds true
  because $\mathcal{N}$ is a grid and $H=1$. Suppose $R_{opt}\geq 2$.
  Consider an optimal rectangulation $\mathcal{R}_{opt}$ and let $H_2$ be the
  number of holes incident with at most 2 segments. As every segment is
  incident with two holes and every hole is incident with at least two
  segments (Lemma \ref{lm:bound_holes_pseudosegments}.(ii)),
  $2H \leq 2S_{opt}$.  Adding $H_2$ respectively $H$ on the sides of this
  inequality yields $3H\leq 2S_{opt}+H_{2}$. In this inequality substitute
  $H_2$ by $4R_{opt}+t(\mathcal{N})$ (Lemma
  \ref{lm:bound_holes_pseudosegments}.(i)) and also substitute $S_{opt}$ by
  $R_{opt}+H-2$ (Lemma \ref{lemma:rectangles_segments_holes}) to obtain
  $ H\leq 6R_{opt}+t(\mathcal{N})-4$. 
\end{proof}

\section{Approximations for the Number of Rectangles}  
\label{sec:approx_rectangles}

The first proposition in this section states that the greedy strategy results
in a 4-approximation when $\delta(\mathcal{N})\geq 3$.  Two families of
strings whose elements have a dual net satisfying this condition are circular
drawings with a bipartite sets of pseudosegments and triangle-free hyperbolic
line arrangements. For references to triangle-free hyperbolic line
arrangements we refer to Eppstein~\cite[Section 7]{Ep09} and his
figure~\cite{Ep-arr} in the Wikipedia article on circle graphs.

\begin{proposition}
If $\delta(\mathcal{N})\geq 3$, then $R_{greed}\leq 4\cdot R_{opt} -2$. 
\end{proposition}
\begin{proof}
Apply Observations \ref{obs:greed_bound_by_rec_holes} and \ref{obs:delta_geq_3}.
\end{proof}

Condition $\delta(\mathcal{N})\geq 3$ is restrictive as it forbids in a set of
strings the existence of a cell bounded by three pieces of strings. The next
lemma handles very general sets of strings at the expense of a larger
approximation factor.

\begin{lemma}\label{lemma:pseudosegments_approx}
  If $\mathcal{N}$ is the dual net of a set of pseudosegments, then
  $R_{greed}\leq 8 R_{opt}+t(\EE)-6$.
\end{lemma}
\begin{proof}
Apply Observation \ref{obs:greed_bound_by_rec_holes} and Lemma \ref{lemma:holes_bounded_by_t-faces}.
\end{proof}

Now our main result is an immediate corollary.
\begin{proof}[Proof of Theorem \ref{thm:main_8approx}]
  Apply the greedy strategy and Lemma \ref{lemma:holes_bounded_by_t-faces} to
  each connected component of the set of strings $\EE$ associated to $D$ to
  obtain the desired bundling.
\end{proof}

\section{Approximating bipartite instances}\label{sec:fivetwo_approx}

A set of strings is \emph{bipartite} if the strings can be colored blue and
red, so that the only crossings are between a blue and a red string. A dual
net $\mathcal{N}$ is bipartite, if it is the net of a bipartite set of
strings. In this section, $\mathcal{N}=(V,E)$ will always be a bipartite
instance. The edges of $\mathcal{N}$ inherit a coloring from the strings when
we color an edge crossed by a red string red and an edge crossed by a blue
string blue.  Note that on the boundary of each square of $\mathcal{N}$ and
around each regular vertex of $\mathcal{N}$ the colors alternate. In
particular every straight-path of $\mathcal{N}$ is either blue or red.

\subsection{A 3/2-approximation for segments in bipartite instances}

Our goal in this section is to approximate the minimum number $S_{opt}$ of
segments needed to rectangulate a bipartite instance.  By first selecting an
initial set of good segments and then applying the greedy strategy to extend
this set into a rectangulation, we will prove the following:

\begin{theorem}\label{thm:fivetwo_approx}
  Let $\mathcal{N}=(V,E)$ be a bipartite dual net. Then there exists a cut-set
  $A$ such that starting the greedy strategy from $A$ produces a saturated
  cut-set $\mathcal{S}_A$ with at most $\frac{3}{2} \cdot S_{opt}$ segments.
\end{theorem}

\begin{definition}[Relative Exponent]
  \marginpar{\raggedright\textit{\small relative exponent}} Given an edge-set
  $E_0\subset E$ of $\mathcal{N}=(V,E)$, the \tem{exponent} of $v$ relative to
  $E_0$, denoted $\text{exp}_{E_0}(v)$, is the minimum number of edges at $v$
  that need to be added to $E_0$ so that the resulting edge-set saturates
  $v$. We let $\text{exp}_{E_0}(\mathcal{N}):=\sum_{v\in V}\text{exp}_{E_0}(v)$. We
  extend this notation to cut-sets~$\mathcal{S}$, so that
  $\text{exp}_\mathcal{S}(v):=\text{exp}_{E_0}(v)$, where $E_0$ is the
  edge-set of the segments in $\mathcal{S}$.
\end{definition}
\begin{remark}
$\text{exp}(v)=\text{exp}_{\emptyset}(v)$.
\end{remark}

We now introduce the \term{marginal gain} of a collection $\mathcal{S}$ of
segments, measuring how good is to use $\mathcal{S}$ as the start of a greedy
strategy.  Let the \tem{marginal gain} of $\mathcal{S}$ be the following
\begin{equation}\label{eq:expg}
g(\mathcal{S})=\sum_{v\in V}(\text{exp}(v)-\text{exp}_\mathcal{S}(v))-|\mathcal{S}|\text{.}\end{equation} 

\begin{observation}\label{obs:gain}
 Suppose that $\mathcal{S}$ is an extension of $\mathcal{S}_0$ obtained from $\mathcal{S}_0$ by applying some steps of
  the greedy strategy.  Then $g(\mathcal{S}_0)\leq g(\mathcal{S})$.  Moreover, if $\mathcal{S}$ induces a
  rectangulation, then $|\mathcal{S}|=\text{exp}(\mathcal{N})-g(\mathcal{S})\leq\text{exp}(\mathcal{N}) -g(\mathcal{S}_0) $. 
\end{observation}

\begin{proof}
  Due to induction, it suffices to consider a 1-step greedy extension
  $\mathcal{S}_1$ of $\mathcal{S}_0$ with $\mathcal{S}_1 = \mathcal{S}_0 +
  s$. Greedy says that there is a vertex $v_1$ such that the exponent at $v$
  is decreased by one upon adding segment $s$ while
  $\text{exp}_{\mathcal{S}_1}(v)\leq \text{exp}_{\mathcal{S}_0}(v)$ for all
  $v$. Therefore,
  $\sum_{v\in V}(\text{exp}(v)-\text{exp}_{\mathcal{S}_1}(v)) \leq \sum_{v\in
    V}(\text{exp}(v)-\text{exp}_{\mathcal{S}_0}(v)) -1$.  Since
  $|\mathcal{S}_1| = |\mathcal{S}_0|+1$ this implies
  $g(\mathcal{S}_1) \geq g(\mathcal{S}_0)$. Finally, the moreover part follows
  from the fact that $\text{exp}_{\mathcal{S}}(\mathcal{N})=0$ when
  $\mathcal{S}$ induces a rectangulation.
\end{proof}

Observation \ref{obs:gain} converts the problem of finding cut-set with
$S_{opt}$ segments into the problem of maximizing $g$: If $\mathcal{S}_{0}^*$
maximizes $g$, then apply the greedy strategy to find a cut-set
$\mathcal{S}^*$ inducing a rectangulation that by Observation \ref{obs:gain}
has $S_{opt}$ segments.  Although we do not know how to compute such an
optimum $\mathcal{S}_0^*$, to prove Theorem \ref{thm:fivetwo_approx} we will
find a set monochromatic set $A$ such that
$g(A)\geq \frac{1}{2}g(\mathcal{S}_0^*)$.

The following considerations are motivated by the aim of finding a minimal set
of blue segments of maximum marginal gain.

Let $B$ be the set of all blue segments in $\mathcal{N}$.
Since a blue segments will never hit another blue segment,
all the segments in $B$ connect two holes, they are the
maximal straight-paths of blue edges. Any set $\mathcal{S}$ of
blue segments is a subset of $B$.

We divide the vertices $v$ of $V$ with $\text{exp}(v)>0$ into two classes: a
vertex $v$ is \term{weak} if $v$ $\text{deg}_\mathcal{N}(v)$ is odd, $v$ is
the boundary of $\mathcal{N}$, and the two boundary edges at $v$ are red. The
rest of the vertices are \term{strong}\footnote{We remark that our definition of weak and strong depends on the color of choice (our current choice is blue). An analogous definition of weak and strong can be made
with respect to the red color.}.  Let $\mathcal{S}\subset B$
is a set of blue segments and let $v$ be a vertex with $\text{exp}(v)>0$ incident to $k$ of the segments.
Since in a bipartite instance every vertex-hole is of even degree the
definition of relative exponent implies that
\begin{equation}\label{strong+weak}
\text{exp}(v)-\text{exp}_\mathcal{S}(v)= \begin{cases}
k & \text{$v$ is strong}\\
\max(k-1,0) & \text{$v$ is weak.}
\end{cases}
\end{equation}

The distinction between weak and strong is helpful to understand the marginal
gain $g$. For instance, if to a set $\mathcal{S}$ we add a segment $s$ between
two strong vertices, then the marginal gain is increased by $1$.  Moreover,
$g(\{s\})>0$ if and only if $s$ connects two strong vertices. So, it would be
natural to think that adding segments between pairs of strong vertices is the
only way to increase $g$. However, we will see that it is possible to add sets
of segments $\mathcal{S}$ where no gain is produced from adding individual
edges from $\mathcal{S}$, but only by adding $\mathcal{S}$ as a whole.

To facilitate the computation of $g$ we consider an auxiliary multigraph
$G_B=(V_B,E_B)$. The vertex-set $V_B$ consist of the vertices of $\mathcal{N}$
with positive exponent. 
The edge-set $E_B$ is the disjoint union $B\cup L$, where $B$ are the blue
segments in $\mathcal{N}$ (each joining its corresponding ends in $V_B$) and
$L$ is a set of loops, one at each strong vertex of $\mathcal{N}$.  The
subgraph of $G_B$ induced by the edges in $\mathcal{S}$ is denoted as
$G_B[\mathcal{S}]$. We ease our notation by letting
$V(\mathcal{S})=V(G_B[\mathcal{S}])$ and
$\text{deg}_\mathcal{S}(v)=\text{deg}_{\mathcal{N}[\mathcal{S}]}(v)$. We
remark that a loop at $v$ contributes $2$ to the degree of $v$.  If
$\mathcal{S}\subseteq B$, then let $\widehat{\mathcal{S}}$ be obtained from
$\mathcal{S}$ by adding all the loops from $L$ at the strong vertices in
$V(\mathcal{S})$.  The next lemma translates $g(\mathcal{S})$ as a function of
$G_B$.

\begin{lemma}\label{lemma:blue_gain}
  Given $\mathcal{S}\subseteq B$, then
  \begin{equation} \label{eq:blue_gain}
    g(\mathcal{S})=
    |\widehat{\mathcal{S}}|-|V(\mathcal{S})|-\text{tc}(\widehat{\mathcal{S}})
  \end{equation}
  where $\text{tc}(\widehat{\mathcal{S}})$ is the number of connected
  components in $G_B[\widehat{\mathcal{S}}]$ isomorphic to a tree.
\end{lemma}  
\begin{proof}
  We proceed by induction on $|\mathcal{S}|$.  Let $f(\mathcal{S})$ be the
  right-hand side of Equation \ref{eq:blue_gain}. For the base case note that
  $g(\emptyset)=0=f(\emptyset)$.  Now consider a non-empty set
  $\mathcal{S}\subseteq B$. By induction assume that
  $g(\mathcal{S}')=f(\mathcal{S}')$ for every proper subset
  $\mathcal{S}'\subset \mathcal{S}$.  Thus, to show
  $g(\mathcal{S})=f(\mathcal{S})$ it is enough to find a proper subset
  $\mathcal{S}'\subset \mathcal{S}$ for which
  $g(\mathcal{S})-g(\mathcal{S}')=f(\mathcal{S})-f(\mathcal{S}')$.

  First, suppose that $\mathcal{S}$ has a subset $C$ such that $G_B[C]$ is a
  connected component in $G_B[\widehat{\mathcal{S}}]$ isomorphic to a cycle of
  length at least $3$, i.e., all the vertices in $V(C)$ are weak.  Then
  $g(\mathcal{S})-g(\mathcal{S}\setminus C)=\sum_{v\in
    V(C)}(\text{exp}_\mathcal{S}(v)-\text{exp}_{\mathcal{S}\setminus
    C}(v))-|C|=\sum_{v\in V(C)}1-|C|=0$. On the other hand, because
  $|C|=|V(C)|$, $f(\mathcal{S})-f(\mathcal{S}\setminus C)=0$. Therefore,
  assuming $g(\mathcal{S}\setminus C)=f(\mathcal{S}\setminus C)$ we obtain
  $g(\mathcal{S})=f(\mathcal{S})$.

  Second, suppose that $G_B[\widehat{\mathcal{S}}]$ has an edge $xy$ for which
  $x$ is a leaf. Let $s\in \mathcal{S}$ be the segment connecting~$x$ and $y$
  and let $\mathcal{S}'=\mathcal{S}-s$. Sinc $x$ has no loop in
  $G_B[\widehat{\mathcal{S}}]$ it is weak, since $s$ it the only segment in
  $S$ incident to $x$ we have
  $\text{exp}_{\mathcal{S}'}(x)-\text{exp}_{\mathcal{S}}(x)=0$. Therefore,
  $g(\mathcal{S})-g(\mathcal{S}')=\text{exp}_{\mathcal{S}'}(y)-\text{exp}_{\mathcal{S}}(y)-1$.
  From Equation~\ref{strong+weak} it follows that
  $g(\mathcal{S})-g(\mathcal{S}')=0$ unless~$y$ is weak and a leaf, in this
  case $g(\mathcal{S})-g(\mathcal{S}')=-1$.  On the other hand
  $f(\mathcal{S})-f(\mathcal{S}')=-\text{tc}(\widehat{\mathcal{S}})+\text{tc}(\widehat{\mathcal{S}'})$.
  Hence, $g(\mathcal{S})-g(\mathcal{S}')$ and $f(\mathcal{S})-f(\mathcal{S}')$
  are equal to $0$ or $-1$ depending on whether $y$ is a leaf in
  $G_B[\widehat{\mathcal{S}}]$ or not.

  Finally, suppose that $G_B[\widehat{\mathcal{S}}]$ neither has leafs nor
  cycle components. In this case, we pick any $s\in \mathcal{S}$ and let
  $\mathcal{S}'=\mathcal{S}-s$. Since the endpoints $x$ and $y$ of $s$ are not
  leafs, $\text{exp}_{\mathcal{S}'}(z)-\text{exp}_{\mathcal{S}}(z)$ is either
  $1$ or $0$ depending whether $z\in \{x,y\}$ or
  $z\in V(\mathcal{S})\setminus \{x,y\}$. This implies that
  $g(\mathcal{S})-g(\mathcal{S}')=1$. Since $G_B[\widehat{\mathcal{S}}]$
  neither has cycle components nor leafs,
  $\text{tc}(\widehat{\mathcal{S}})=0=\text{tc}(\widehat{\mathcal{S}'})$ and
  $V(\mathcal{S})=V(\mathcal{S}')$. Therefore
  $f(\mathcal{S})-f(\mathcal{S}')=1$, and we are done.
\end{proof}

With the help of Lemma \ref{lemma:blue_gain} we can now build examples of sets
$\mathcal{S}\subseteq B$ with $g(\mathcal{S})>0$ but where segments in
$\mathcal{S}$ are not necessarily between strong vertices. A \term{bicycle} is
a connected graph with no degree-1 vertices that has exactly two cycles. Every
bicycle is one of three kinds: a \tem{theta-graph}, obtained from a cycle by
adding a path connecting two vertices of the cycle; an \tem{eight-graph},
obtain by gluing two disjoint cycles at exactly one vertex; or a
\tem{barbell}, obtained by joining two disjoint cycles by a path.

Note that
$|\widehat{\mathcal{S}}| = \sum_v\text{deg}_{\widehat{\mathcal{S}}}(v)/2 =
\sum_v\text{deg}_{\mathcal{S}}(v)/2 + \#(\text{strong vertices in
  $V(\mathcal{S})$})$. If $\mathcal{S}$ induces a bicycle in $G_B$, then
$|\mathcal{S}|=\frac{1}{2}\sum_v\text{deg}_{\mathcal{S}}(v) = |V(\mathcal{S})|
+1$, therefore Eq. \ref{eq:blue_gain} implies $g(\mathcal{S})>0$.

Bicycles are somehow related to a matroid and although we will not use Matroid Theory here, some of the  concepts that we define next are well known in this context.

A \tem{pseudoforest} is a graph where each connected component has at most one
cycle.  Given the graph $G_B=(V_B,E_B)$, its \term{bicircular matroid} is the pair $M=(E_B, \mathcal{I})$ where $\mathcal{I}$ is the set of subsets of $E_B$ inducing a pseudoforest in $G_B$. In the matroid context, the subsets of $E_B$ in $\mathcal{I}$ are called {\em independent} while the subsets not in $\mathcal{I}$  are called {\em dependent}. Indeed, the bicycles of $G_B$ are the minimal dependent
sets of $M$, known as circuits. The {\em rank function} $\text{rk}(\cdot):2^{E_B}\rightarrow \mathbb{Z}_{\geq 0}$ assigns to each $S\subseteq E$ the size $\text{rk}(S)$ of a maximum independent set contained
in $S$. The \tem{nullity} of $S$ is defined as
$\text{null}(S)=|S|-\text{rk}(S)$.

We will use two basic
properties of bicircular matroids whose proofs are left to the reader: (P1)
$\text{rk}(\mathcal{S})=|V(\mathcal{S})|-\text{tc}(\mathcal{S})$ where
$\text{tc}(\mathcal{S})$ is the number of tree components of
 $G_B[\widehat{\mathcal{S}}]$; and (P2)
$\text{null}(\mathcal{S}_1)\leq \text{null}(\mathcal{S}_2)$ for
$\mathcal{S}_1\subseteq \mathcal{S}_2$. The next
observation follows from Lemma \ref{lemma:blue_gain} and P1:

\begin{observation}\label{obs:g-null}
  For $\mathcal{S}\subseteq B$,
  $g(\mathcal{S})=\textrm{null}(\widehat{\mathcal{S}})-2\cdot
  \text{tc}(\widehat{\mathcal{S}})$.\qed
\end{observation}

 We are now ready for finding a blue set  maximizing $g$:

\begin{lemma}\label{lemma:maxblue}
  There is a unique set $A\subseteq B$ such that 
   for all $\mathcal{S}\subset B$,
  $g(\mathcal{S})\leq g(A)$. 
  Moreover, $A$ can be obtained in $O(|E_B|)$ time.
\end{lemma}

\begin{proof}
  We start by pruning $E_B$, that is, we iteratively remove edges incident
  with leaves until no leaf remains. Afterwards, we remove all the connected
  components isomorphic to cycles. The resulting edge-set $A$ is the one we
  are looking for.

  Note that P1 implies that the nullity of a set set does not
  change after removing a cycle component, or after pruning a leaf, this also
  holds when pruning the only edge of a tree. Thus,
  $g(A)=\text{null}(A)-0=\text{null}(E_B)$. Now Observation~\ref{obs:g-null}
  and P2 imply that for $\mathcal{S}\subseteq B$,
  $g(\mathcal{S})\leq \text{null}(\mathcal{S})\leq
  \text{null}(E_B)=g(A)$.
  \end{proof}
    
  We now turn our head back to the proof of Theorem \ref{thm:fivetwo_approx}
  where also red segments are considered.  Analogous to $B$, we let $R$ be the
  set of red segments connecting vertices with positive exponent.

  Given a saturating set of segments $\mathcal{S}$ (not necessarily
  monochromatic), a \term{blue-red split} is a partition $\mathcal{S}_B$,
  $\mathcal{S}_R$ of $\mathcal{S}$, where $\mathcal{S}_B$ are the blue segments
  and $\mathcal{S}_R$ are the red segments. Note that the straight-paths in
  $\mathcal{S}_B$ or in $\mathcal{S}_R$ do not necessarily have ends in holes.

\begin{lemma}\label{lemma:split}
  Let $\mathcal{S}_B$, $\mathcal{S}_R$ be a blue-red split of a saturating
  cut-set $\mathcal{S}$. Then
$$g(\mathcal{S}_B)+g(\mathcal{S}_R)\geq g(\mathcal{S})\text{.}$$
\end{lemma}
\begin{proof}
  From the definition of $g(\cdot)$ (Eq. \ref{eq:expg}) and since
  $|\mathcal{S}_B|+|\mathcal{S}_R|= |\mathcal{S}|$ it is enough to show that
  for every $v$ with $\text{exp}(v)>0$,

\begin{equation}\label{eq:equivalent}
  \text{exp}_\mathcal{S}(v)+\text{exp}(v)\geq
          \text{exp}_{\mathcal{S}_B}(v)+\text{exp}_{\mathcal{S}_R}(v)\text{.} 
\end{equation}
Since $\mathcal{S}$ is saturating we have $\text{exp}_{\mathcal{S}}(v)=0$.  Let
$\mathcal{S}_R(v)$, $\mathcal{S}_B(v)$, and $\mathcal{S}(v)$ denote the
segments in $\mathcal{S}_R$, $\mathcal{S}_B$, and $\mathcal{S}$ respectively
which are incident to $v$. From the definition of the sets,
$|\mathcal{S}_R(v)|+ |\mathcal{S}_B(v)| =|\mathcal{S}(v)|$.

If $v$ is strong in both colors, then from Eq.~\ref{strong+weak} we know
$\text{exp}_{\mathcal{S}_B}(v) = \text{exp}(v) - |\mathcal{S}_B(v)|$ and
$\text{exp}_{\mathcal{S}_R}(v) = \text{exp}(v) - |\mathcal{S}_R(v)|$. Hence,
Eq.~\ref{eq:equivalent} reduces to
$|\mathcal{S}_R(v)|+ |\mathcal{S}_B(v)| \geq \text{exp}(v)$ which is true
because $|\mathcal{S}_R(v)|+ |\mathcal{S}_B(v)| =|\mathcal{S}(v)|$ and
$\mathcal{S}$ is saturating.

Now let $v$ be weak in one of the colors and note that it is strong in the
other.  By symmetry we may assume that $v$ is weak in red. If
$\mathcal{S}_R(v)=\emptyset$ we have
$\text{exp}_{\mathcal{S}_R}(v) = \text{exp}(v) - |\mathcal{S}_R(v)|$ as
above. If $|\mathcal{S}_R(v)| > 0$, then
$\text{exp}_{\mathcal{S}_R}(v) = \text{exp}(v) - |\mathcal{S}_R(v)| +1$ and we
have to verify
$|\mathcal{S}(v)| = |\mathcal{S}_R(v)|+ |\mathcal{S}_B(v)| \geq
\text{exp}(v)+1$. For a vertex of the given kind, however, the unique
saturating set of edges of size $\text{exp}(v)$ is the set of all blue edges.
Since $\mathcal{S}$ is saturating and contains a red segment at $v$ the
required inequality $|\mathcal{S}(v)| \geq \text{exp}(v)+1$ holds.
\end{proof}
  
\begin{proof}[Proof of Theorem \ref{thm:fivetwo_approx}]
  Let $\mathcal{S}$ be a saturating set of segments of minimum cardinality.
  Let $\mathcal{S}_B$, $\mathcal{S}_R$ be the blue-red split of $\mathcal{S}$.
  By possibly interchanging colors, we may assume that
  $g(\mathcal{S}_B)\geq g(\mathcal{S}_R)$. Lemma \ref{lemma:split} implies
  that $g(\mathcal{S}_B)\geq \frac{1}{2} g(\mathcal{S})$.

  Consider a set $\mathcal{S}_B'\subseteq B$, obtained by extending the
  straight-paths of $\mathcal{S}_B$ into elements of $B$. A segment of
  $\mathcal{S}_B'$ may contain multiple elements of $\mathcal{S}_B$, and also
  the exponent of each vertex relative to $\mathcal{S}_B'$ is at most its
  exponent relative to $\mathcal{S}_B$. Thus,
  $g(\mathcal{S}_B')\geq g(\mathcal{S}_B)$.

  Let $A\subset B$ be the unique minimal set of blue segments maximizing the
  gain, see Lemma~\ref{lemma:maxblue}, and let $A^+$ be a saturating extension
  of $A$ which is obtained by the greedy strategy. Then Obs.~\ref{obs:gain}
  and the properties of $A$ imply the first and second inequalities in
  $g(A^+)\geq g(A)\geq g(\mathcal{S}_B')\geq g(\mathcal{S}_B)\geq
  \frac{1}{2}g(\mathcal{S})$.

  Since $\mathcal{S}$ and $A^+$ are saturating
  $|\mathcal{S}|=\text{exp}(\mathcal{N})-g(\mathcal{S})$, and
  $|A^+|=\text{exp}(\mathcal{N})-g(A^+)$.  {} From the above
$$
|A^+| = \text{exp}(\mathcal{N})-g(A^+)\leq
\text{exp}(\mathcal{N})-\frac{1}{2}g(\mathcal{S})=\frac{1}{2}\text{exp}(\mathcal{N})
+ \frac{1}{2}|\mathcal{S}|
$$
Finally, Obs. \ref{obs:approx_seg}.\ref{it:seg_opt} tells us that
$\text{exp}(\mathcal{N})/2\leq |\mathcal{S}|$ which yields
$|A^+|\leq \frac{3}{2}|\mathcal{S}|$.
\end{proof}  

\subsection{A 9/2-approximation for rectangles in bipartite instances}

\begin{proof}[Proof of Theorem \ref{thm:approx_bipartite}]
  By Theorem \ref{thm:fivetwo_approx}, we can compute a rectangulation
  $\mathcal{R}$ with $R$ rectangles, $S$ segments, and
  $S\leq \frac{3}{2}S_{opt}$. Substitute both sides of this inequality using
  Eq. \ref{eq:rsh} to obtain $R\leq \frac{3}{2}R_{opt}
  +\frac{1}{2}H-1$. Substitute $H$ by $6R_{opt}+t-4$ (Lemma
  \ref{lemma:holes_bounded_by_t-faces}) to obtain
  $R\leq \frac{9}{2}R_{opt}+\frac{1}{2}t-3\leq
  \frac{9}{2}R_{opt}+\frac{1}{2}t$.
\end{proof}

\section{Conclusion}

In this paper we studied the bundled crossing number of connected good
drawings and showed that the greedy strategy derived from the problem of
rectangulating an orthogonal polygon leads to an 8-approximation (up to adding
the number of toothed-faces in the drawing). Moreover, we improved this strategy
for bipartite instances by considering an initial good set of segments. We
hope that the tools and the framework developed in this work will inspire more
results about bundled crossings. We leave below some We leave below some open questions.

\begin{enumerate}
\item Is there a constant $c$ guaranteeing that, for any simple drawing $D$,
  the greedy algorithm produces bundling with at most $c\cdot \text{bc}(D)$
  bundles? In other words, are toothed-faces relevant to approximate
  $\text{bc}(D)$?
\item What is the computational complexity of computing $\text{bc}(\cdot)$ for
  bipartite instances?
\end{enumerate}

\bibliography{bundles.bib}

\begin{thebibliography}{10}

\bibitem{AFP16}
{\sc M.~J. Alam, M.~Fink, and S.~Pupyrev}, {\em The bundled crossing number},
  in Proc. {GD} 2016, vol.~9801 of LNCS, Springer, 2016, pp.~399--412.

\bibitem{BRHMD17}
{\sc B.~Bach, N.~H. Riche, C.~Hurter, K.~Marriott, and T.~Dwyer}, {\em Towards
  unambiguous edge bundling: Investigating confluent drawings for network
  visualization}, {IEEE} Trans. Vis. Comput. Graph., 23 (2017), 541--550.

\bibitem{beineke2010early}
{\sc L.~Beineke and R.~Wilson}, {\em The early history of the brick factory
  problem}, The Mathematical Intelligencer, 32 (2010), 41--48.

\bibitem{CD+19}
{\sc S.~Chaplick, T.~C. van Dijk, M.~Kryven, J.~Park, A.~Ravsky, and A.~Wolff},
  {\em Bundled crossings revisited}, in Proc. GD 2019, vol.~11904 of LNCS,
  Springer, 2019, pp.~63--77.

\bibitem{CuiZQWL08}
{\sc W.~Cui, H.~Zhou, H.~Qu, P.~C. Wong, and X.~Li}, {\em Geometry-based edge
  clustering for graph visualization}, {IEEE} Trans. Vis. Comput. Graph., 14
  (2008), 1277--1284.

\bibitem{DEGM05}
{\sc M.~Dickerson, D.~Eppstein, M.~T. Goodrich, and J.~Y. Meng}, {\em Confluent
  drawings: Visualizing non-planar diagrams in a planar way}, J. Graph
  Algorithms Appl., 9 (2005), 31--52.

\bibitem{Ep-arr}
{\sc D.~Eppstein}.
\newblock \url{https://en.wikipedia.org/wiki/Circle_graph#/media/File:}.

\bibitem{Ep09}
{\sc D.~Eppstein}, {\em Graph-theoretic solutions to computational geometry
  problems}, in Proc {WG} 2009, vol.~5911 of LNCS, 2009, pp.~1--16.

\bibitem{EppGMHN05}
{\sc D.~Eppstein, M.~T. Goodrich, and J.~Y. Meng}, {\em Delta-confluent
  drawings}, in Proc. {{GD}}'05, vol.~3843 of {{LNCS}}, {Springer}, 2005,
  pp.~165--176.

\bibitem{EppHLNSV16}
{\sc D.~Eppstein, D.~Holten, M.~L{\"{o}}ffler, M.~N{\"{o}}llenburg,
  B.~Speckmann, and K.~Verbeek}, {\em Strict confluent drawing}, J. of Comp.
  Geom. (JoCG), 7 (2016), 22--46.

\bibitem{EppsteinS13}
{\sc D.~Eppstein and J.~A. Simons}, {\em Confluent {H}asse diagrams}, J. Graph
  Algorithms Appl., 17 (2013), 689--710.

\bibitem{ErsoyHPCT11}
{\sc O.~Ersoy, C.~Hurter, F.~V. Paulovich, G.~Cantareiro, and A.~Telea}, {\em
  Skeleton-based edge bundling for graph visualization}, {IEEE} Trans. Vis.
  Comput. Graph., 17 (2011), 2364--2373.

\bibitem{FHSV16}
{\sc M.~Fink, J.~Hershberger, S.~Suri, and K.~Verbeek}, {\em Bundled crossings
  in embedded graphs}, in Proc. {LATIN} 2016, vol.~9644 of LNCS, Springer,
  2016, pp.~454--468.

\bibitem{FinkPW15}
{\sc M.~Fink, S.~Pupyrev, and A.~Wolff}, {\em Ordering metro lines by block
  crossings}, J. Graph Algorithms Appl., 19 (2015), 111--153.

\bibitem{FGKN19}
{\sc H.~F{\"{o}}rster, R.~Ganian, F.~Klute, and M.~N{\"{o}}llenburg}, {\em On
  strict (outer-)confluent graphs}, in Proc. {{GD}}'19, vol.~11904 of {{LNCS}},
  {Springer}, 2019, pp.~147--161.

\bibitem{GansnerHNS11}
{\sc E.~R. Gansner, Y.~Hu, S.~C. North, and C.~E. Scheidegger}, {\em Multilevel
  agglomerative edge bundling for visualizing large graphs}, in Proc.
  PacificVis'11, {IEEE}, 2011, pp.~187--194.

\bibitem{GK06}
{\sc E.~R. Gansner and Y.~Koren}, {\em Improved circular layouts}, in Proc.
  {{GD}}'06, vol.~4372 of {{LNCS}}, {Springer}, 2006, pp.~386--398.

\bibitem{harary1963number}
{\sc F.~Harary and A.~Hill}, {\em On the number of crossings in a complete
  graph}, Proceedings of the Edinburgh Mathematical Society, 13 (1963),
  333--338.

\bibitem{harary1973toroidal}
{\sc F.~Harary, P.~C. Kainen, and A.~J. Schwenk}, {\em Toroidal graphs with
  arbitrarily high crossing numbers}, Nanta Math, 6 (1973), 58--67.

\bibitem{Holten06}
{\sc D.~Holten}, {\em Hierarchical edge bundles: Visualization of adjacency
  relations in hierarchical data}, {IEEE} Trans. Vis. Comput. Graph., 12
  (2006), 741--748.

\bibitem{HoltenW09}
{\sc D.~Holten and J.~J. van Wijk}, {\em Force-directed edge bundling for graph
  visualization}, Comput. Graph. Forum, 28 (2009), 983--990.

\bibitem{HuiPSS07}
{\sc P.~Hui, M.~J. Pelsmajer, M.~Schaefer, and D.~Stefankovic}, {\em Train
  tracks and confluent drawings}, Algorithmica, 47 (2007), 465--479.

\bibitem{LambertBA10}
{\sc A.~Lambert, R.~Bourqui, and D.~Auber}, {\em Winding roads: Routing edges
  into bundles}, Comput. Graph. Forum, 29 (2010), 853--862.

\bibitem{PupyrevNBH16}
{\sc S.~Pupyrev, L.~Nachmanson, S.~Bereg, and A.~E. Holroyd}, {\em Edge routing
  with ordered bundles}, Comput. Geom., 52 (2016), 18--33.

\bibitem{survey1}
{\sc M.~{Schaefer}}, {\em {The graph crossing number and its variants: a
  survey.}}, {Electron. J. Comb.}, Dynamic Survey 21 (2013), p.~90.

\bibitem{turan1977welcome}
{\sc P.~Turán}, {\em A note of welcome}, Journal of Graph Theory, 1 (1977),
  7--9.

\bibitem{DF+17}
{\sc T.~C. van Dijk, M.~Fink, N.~Fischer, F.~Lipp, P.~Markfelder, A.~Ravsky,
  S.~Suri, and A.~Wolff}, {\em Block crossings in storyline visualizations}, J.
  Graph Algorithms Appl., 21 (2017), 873--913.

\bibitem{ZPG19}
{\sc J.~X. Zheng, S.~Pawar, and D.~F.~M. Goodman}, {\em Further towards
  unambiguous edge bundling: Investigating power-confluent drawings for network
  visualization}, IEEE Transactions on Visualization and Computer Graphics,
  (2019).

\end{thebibliography}
\bibliographystyle{my-siam}

\end{document}